\documentclass[11pt]{article}
\usepackage{amsmath,amsfonts,amssymb,amsthm}
\usepackage{graphicx}
\usepackage{subcaption}
\usepackage{array}
\usepackage{supertabular}
\usepackage{bm}
\usepackage{cases}

\begin{document}

\title{Convergence of Estimative Density to Information Projection for Misspecified Normal Distribution Model}
\author{Yo Sheena}
\date{\today}
\maketitle

\section{Introduction}
Consider the $k$-dimensional multivariate normal model $N_k(\mu,\Sigma)$, whose density is given by
\begin{equation}
g(x;\theta) (\triangleq g(x;\mu, \Sigma)) = \frac{|\Sigma^{-1}|^{1/2}}{(2\pi)^{k/2} } \exp\{-\frac{1}{2}(x-\mu)^t \Sigma^{-1} (x-\mu)\},
\end{equation}
where $\mu (\in R^k)$ is the mean vector and $\Sigma $ is a $k$-dimensional positive-definite covariance matrix.

The canonical representation of $g(x; \theta)=g(x; \mu, \Sigma)$ as an exponential family is given by
\[
g(x; \theta) = \exp\left(\sum_{i=1}^k \theta_i \xi_i(x) + \sum_{i \leq j} \theta_{ij} \xi_{ij}(x)- \Psi(\theta) \right),
\]
where 
\begin{align*}
&\xi(x) = (\xi_M(x), \xi_V(x)),\quad \xi_M(x)= (\xi_1(x),\ldots,\xi_k(x)),\\
&\xi_V(x) = (\xi_{11}(x),\xi_{12}(x),\ldots,\xi_{1k}(x),\xi_{22}(x),\xi_{23}(x),\ldots,\xi_{2k}(x),\ldots,\xi_{kk}(x)),\\
&\xi_i(x) = x_i \ (1\leq i \leq k),\quad \xi_{ij}(x) = x_i x_j \ (1 \leq i \leq j \leq k),\\
&\theta = (\theta_M, \theta_V), \quad \theta_M = (\theta_1,\ldots,\theta_k),\quad \theta_V= (\theta_{11},\theta_{12},\ldots,\theta_{1k},\theta_{22}, \theta_{23},\ldots \theta_{2k},\ldots, \theta_{kk} ),\\
& \theta_i = \sum_{i=j}^k \mu_j \sigma^{ij}\ (1\leq i \leq k),\qquad
\theta_{ij}= 
\begin{cases}
-\frac{1}{2} \sigma^{ij} & \text{if $1\leq i=j \leq k$,}\\
-\sigma^{ij} & \text{if $1\leq i < j \leq k$}.
\end{cases}
\end{align*}
The dimension of the model is $p=k(k+3)/2$.

We assume that the true distribution is the $k$-dimensional multivariate $t$ distribution $t_k(0, I_k, \nu)$; that is, its location vector, scale matrix, and degrees of freedom are the $k$-dimensional zero vector, the $k$-dimensional identity matrix, and $\nu$, respectively. We assume $\nu > 6$ to ensure that the required moment conditions hold. The density function $g(x)$ is given by
\begin{equation}
\label{t-dist density}
g(x) = C(\nu) \biggl\{1+\frac{1}{\nu}\sum_{i=1}^k x_i^2 \biggr\}^{-(\nu+k)/2},\qquad C(\nu) = \frac{ \Gamma((\nu+k)/2)}{\Gamma(\nu/2)(\pi\nu)^{k/2}}
\end{equation}

In a more general framework, we define the ``information projection" (see Csiszár \cite{Csiszar1}).
Consider a parametric family of probability distributions on a space $\mathfrak{X}$, denoted by $\mathcal{M}$, consisting of positive-valued densities $g(x;\theta)$ with respect to a measure $\mu$:
\begin{equation}
\label{model_M}
\mathcal{M} = \{ g(x;\theta) \mid \theta = (\theta^1,\ldots,\theta^p) , \theta \in \Theta \},
\end{equation}
where $\Theta$ is an open subset of $R^p$, and $g(x;\theta_1) = g(x;\theta_2)$ almost everywhere if and only if $\theta_1=\theta_2$.
Let $D[\cdot | \cdot]$ denote the Kullback--Leibler divergence (K--L divergence), and let $g(x)$ denote the density of the true distribution. Define $\theta_*$ by
\begin{equation}
\label{info_project}
\theta_* = \underset{\theta \in \Theta}{argmin} D[g(x) | g(x;\theta)].
\end{equation}
The density $g(x;\theta_*)$ is called the information projection. In other words, the information projection is the ``nearest" point in $\mathcal{M}$ to the true distribution.

We first determine the information projection in the present setting, where $\mathcal{M}$ is the family of $k$-dimensional normal distributions and the true distribution is $t_k(0, I_k, \nu)$.
\begin{align}
&\underset{\mu, \Sigma^{-1}}{argmin} D[g(x) | g(x;\mu,\Sigma)] \nonumber\\
&=\underset{\mu, \Sigma^{-1}}{argmin} \int_{R^k} g(x)\{ \log g(x) -\log g(x;\mu, \Sigma)\} dx \nonumber\\
&=\underset{\mu, \Sigma^{-1}}{argmax} \int_{R^k} g(x)\log g(x;\mu, \Sigma)\} dx \nonumber\\
&=\underset{\mu, \Sigma^{-1}}{argmax} \int_{R^k} g(x) \left\{\frac{1}{2} \log |\Sigma^{-1}| -\frac{1}{2}(x-\mu)^t\Sigma^{-1}(x-\mu)\right\} dx \nonumber\\
&=\underset{\mu, \Sigma^{-1}}{argmin} \int_{R^k} g(x) \left\{ -\log |\Sigma^{-1}| + (x-\mu)^t\Sigma^{-1}(x-\mu)\right\} dx \label{int_D}
\end{align}
Since 
\begin{align*}
& \frac{\partial }{\partial \mu} \int_{R^k} g(x) \left\{ -\log |\Sigma^{-1}| + (x-\mu)^t\Sigma^{-1}(x-\mu)\right\} dx \\
&= 2 \Sigma^{-1} \int_{R^k} g(x) (\mu-x) dx =2 \Sigma^{-1}\mu
\end{align*}
we find that
\[
\int_{R^k} g(x) \left\{ -\log |\Sigma^{-1}| + (x-\mu)^t\Sigma^{-1}(x-\mu)\right\} dx
\]
is minimized at $\mu=0$. We next evaluate the integral
\begin{equation}
\label{derive_info_pro}
\int_{R^k} g(x) \left\{ -\log |\Sigma^{-1}| + x^t\Sigma^{-1}x\right\} dx =  -\log |\Sigma^{-1}| + \mathrm{tr} \Sigma^{-1}\int_{R^k} g(x) x^t x\; dx
\end{equation}
If $x \sim t_k(0, I_k, \nu)$, then $E[x^t x] = \frac{\nu}{\nu-2} I_k$. Thus, \eqref{derive_info_pro} becomes
\[
-\log |\Sigma^{-1}| +\frac{\nu}{\nu-2} \mathrm{tr} \Sigma^{-1}
\]
The partial derivative of this expression with respect to $\sigma^{ij} \triangleq (\Sigma^{-1})_{ij}$ is
\[
- \sigma_{ij} + \frac{\nu}{\nu-2} \text{ if $i = j$}, \qquad 
-2 \sigma_{ij} \text{ if $i \ne j$}.
\]
Therefore
\[
\int_{R^k} g(x) \left\{ -\log |\Sigma^{-1}| + x^t\Sigma^{-1}x\right\} dx
\]
is minimized at $\Sigma=\frac{\nu}{\nu-2} I_k$. Consequently, the information projection is
\[
g(x;\theta_*) = g(x; \mu_*, {\Sigma_*}),\qquad \mu_* = 0, \quad \Sigma_*=\lambda_1 I_k, \quad \lambda_1 \triangleq \frac{\nu}{\nu-2}.
\]

We measure the discrepancy between the information projection $g(x;\theta_*)$ and the estimated density $g(x;\hat{\theta})$, where $\hat{\theta}$ is the maximum likelihood estimator, using the risk
\begin{equation}
\label{risk_ch2}
R[g(x;\theta_*)\,|\,g(x;\hat{\theta})] \triangleq E[D[g(x;\theta_*)\,|\,g(x;\hat{\theta})]].
\end{equation}
Here, $E[\cdot]$ denotes expectation with respect to the true distribution $g(x)$.

A general asymptotic expansion of the risk is given in \cite{Sheena4}. Applying Theorem 2 of \cite{Sheena4} to the present problem yields an asymptotic approximation to $R[g(x;\theta_*) | g(x;\hat{\theta})]$. We state only the result here; the detailed calculations are presented in the subsequent sections.

The second-order approximation to the risk, denoted by $f(n, k, \nu)$, is given by
\begin{equation}
\label{second-order_approximation}
\begin{aligned}
&f(n, k, \nu)\\
&=\frac{1}{2n}\Big(p+\frac{k(k+2)}{\nu-4}\Big)\\
&\qquad
+\frac{k}{24n^2(\nu-6)(\nu-4)^2}\\
&\qquad\quad\times
\Big[
(10k^2+63k+49)\nu^3
-(116k^2+858k+590)\nu^2\\
&\qquad\qquad\quad
+(344k^2+3276k+1800)\nu
-(304k^2+3912k+2192)
\Big],
\end{aligned}
\end{equation}
while the first-order approximation is given by
\[
\tilde{f}(n, k, \nu) \triangleq \frac{1}{2n}\Big(p+\frac{k(k+2)}{\nu-4}\Big).
\]

We first compare these two approximations with the simulated risk. The simulation is performed by generating $n$ observations from $t_k(0, I_k, \nu)$ and then calculating the MLE $\hat{\theta}$ and the K--L divergence $D[g(x;\theta_*) | g(x;\hat{\theta})]$. We repeat this procedure $N$ times and average the resulting $N$ values of $D[g(x;\theta_*) | g(x;\hat{\theta})]$.

\begin{figure}[t]
\centering
\begin{subfigure}{0.48\textwidth}
  \centering
  \includegraphics[width=\linewidth]{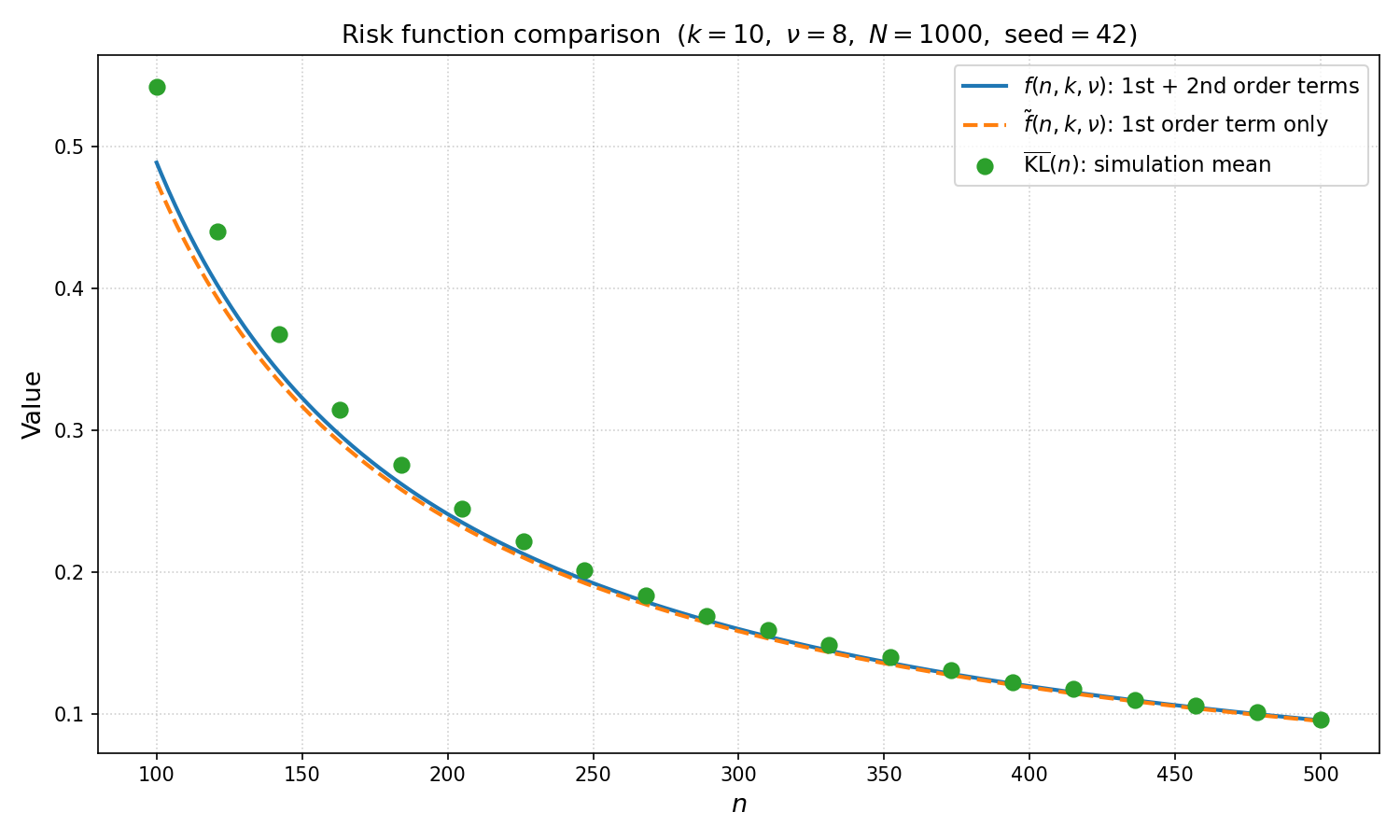}
  \label{fig:risk-k10-nu8}
\end{subfigure}
\hfill
\begin{subfigure}{0.48\textwidth}
  \centering
  \includegraphics[width=\linewidth]{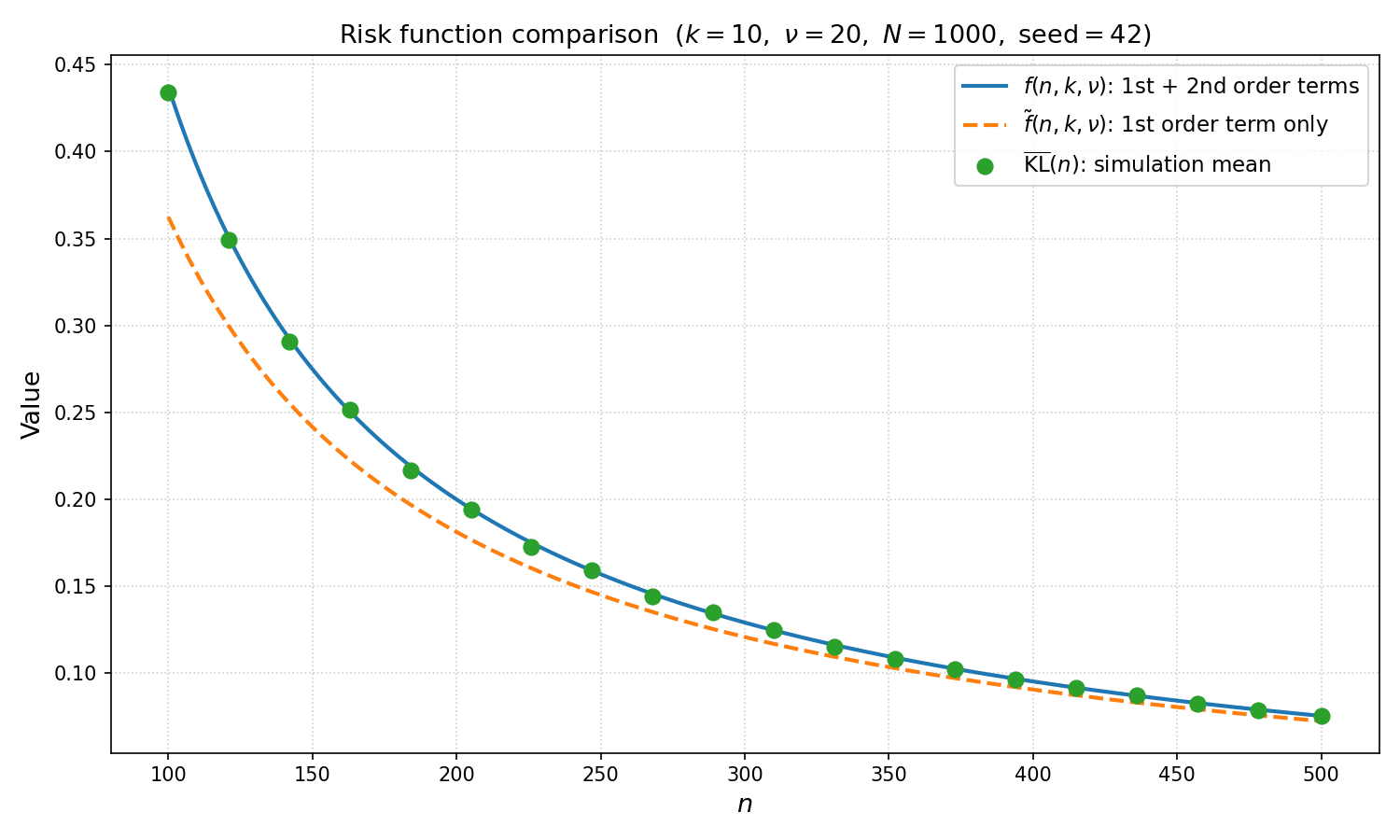}
  \label{fig:risk-k10-nu20}
\end{subfigure}

\vspace{3mm}

\begin{subfigure}{0.48\textwidth}
  \centering
  \includegraphics[width=\linewidth]{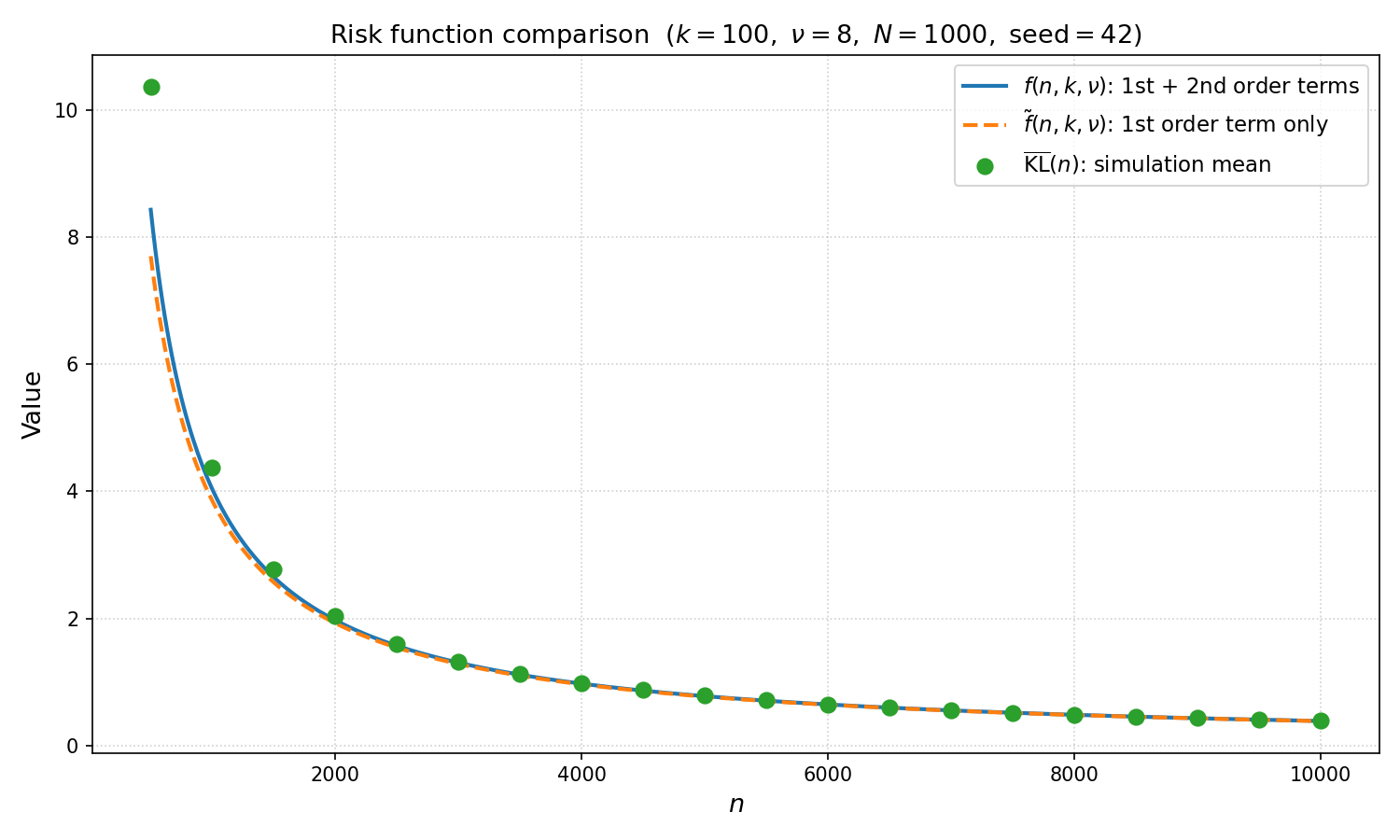}
  \label{fig:risk-k100-nu8}
\end{subfigure}
\hfill
\begin{subfigure}{0.48\textwidth}
  \centering
  \includegraphics[width=\linewidth]{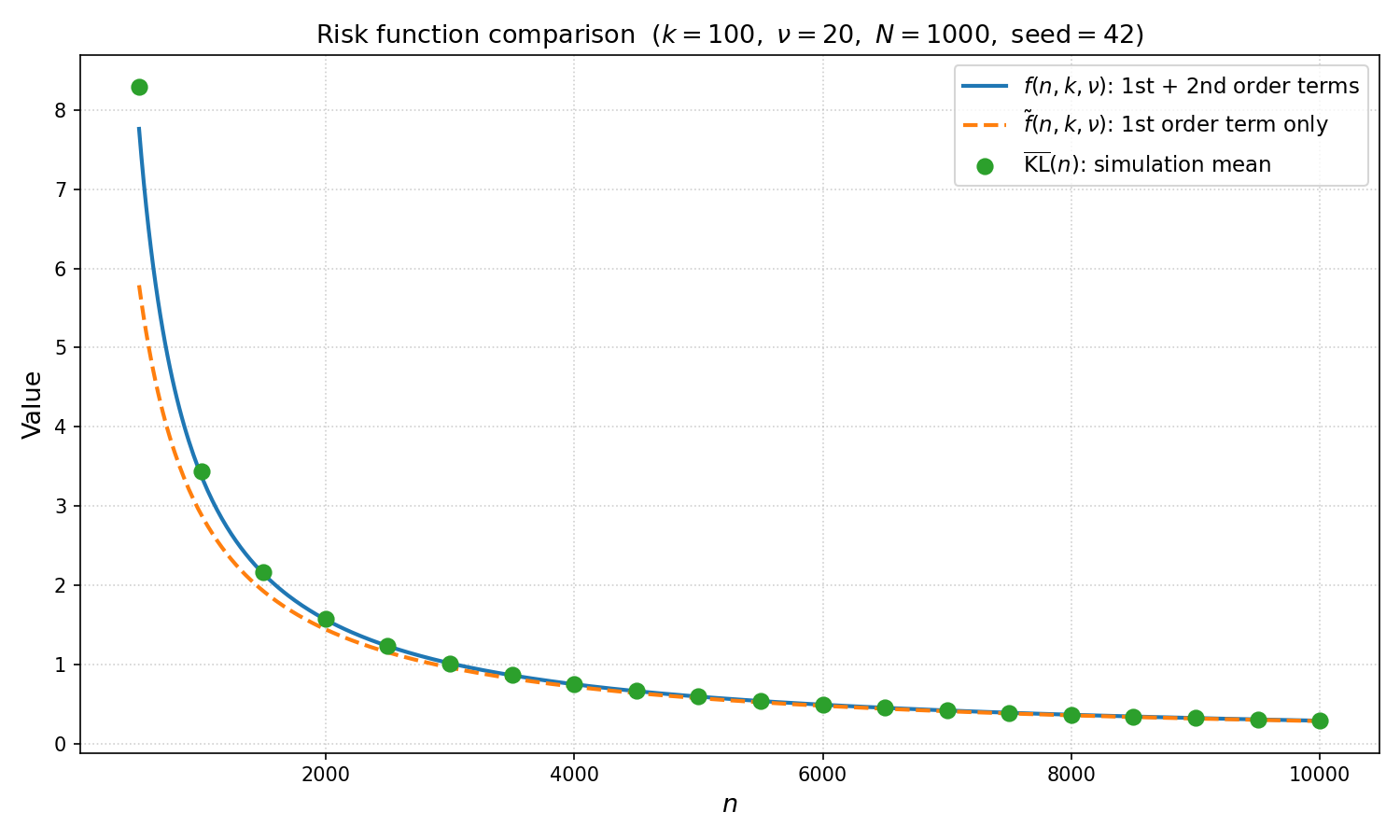}
  \label{fig:risk-k100-nu20}
\end{subfigure}

\caption{Comparison of the risk functions.}
\label{three_risks_t_dist_normal}
\end{figure}

In Figure \ref{three_risks_t_dist_normal}, the three risk functions of $n$, namely $f(n,k,\nu)$, $\tilde{f}(n,k,\nu)$, and the simulated risk $\overline{KL}(n)$, are plotted with the $x$-axis representing $n$ for the four combinations $(k,\nu)=(10, 8), (10, 20), (100, 8), (100,20)$, with $N$ fixed at 1,000.

In Chapter 3 of \cite{Sheena4}, we derived the threshold $C_\alpha$ that guarantees sufficient closeness between two distributions in terms of the Bayes error rate. Equation (17) in \cite{Sheena4} gives the threshold $C_\alpha$ for a prescribed Bayes error rate $1/2-\alpha$. Setting $\alpha= 0.05$, the value of $C_{0.05}$ is $1/50$. For given $k$ and $\nu$, either approximation, $f(n, k, \nu)$ or $\tilde{f}(n, k, \nu)$, can be used together with this threshold to determine $n^*$ such that
\[
n^*=\min \{\, n \in \mathbb{Z} \mid \tilde{f}(n,k,\nu) \leq 1/50 \,\}.
\]
or
\[
n^*=\min \{\, n \in \mathbb{Z} \mid f(n,k,\nu) \leq 1/50 \,\}.
\]
We again consider the four cases $(k,\nu)=(10, 8), (10, 20), (100, 8), (100,20)$. Table \ref{n_search_for_t_dist} reports the values of $n^*$ obtained from $\tilde{f}$ and $f$, together with the value of $n^*$ obtained from the simulated risk $\overline{KL}$.
\begin{table}[htbp]
\centering
\caption{Sample size required for $\alpha=0.05$}
\label{n_search_for_t_dist}
\begin{tabular}{c|c|c|c}
                      & $\tilde{f}$ & $f$  & $\overline{KL}$ \\
\hline
$(k,\nu)=(10,8)$      & 2375        & 2378 & 2394       \\
\hline
$(k,\nu)=(10,20)$     & 1813        & 1833 & 1827       \\
\hline
$(k,\nu)=(100,8)$     & 192501      & 192548 & 192593    \\
\hline
$(k,\nu)=(100,20)$    & 144688      & 144858 & 144910    \\
\hline
\end{tabular}
\end{table}
%
%
%
%
\section{Basic Moments}
Before calculating the various moments appearing in Theorem 2 of \cite{Sheena4}, we present basic formulas for the moments of $x=(x_1,\ldots,x_k)$ when
\[
x \sim N_k(0, \lambda_1I_k),
\]
or
\[
x \sim t_k(0, I_k, \nu).
\]

For calculating higher-order moments of $N_k(0, \lambda_1 I_k)$, the following formula is useful. For $1\leq a_1, a_2, \ldots, a_n \leq k$,
\begin{equation}
\label{Wick_formula}
E[x_{a_1} x_{a_2}\cdots x_{a_n}] =
\begin{cases}
\lambda_1^{n/2}\sum_{p \in P} \prod_{\{i,j\} \in p} \delta_{ij} & \text{ if $n$ is even,}\\
0 & \text{ if $n$ is odd,}
\end{cases}  
\end{equation}
where $P$ is the set of all pairings of $\{a_1, \ldots, a_n\}$, that is, all distinct partitions of $\{a_1, \ldots, a_n\}$ into $n/2$ pairs, and the product is taken over the pairs $\{i, j\}$ in a pairing $p$.

The moments for $x \sim  t_k(0, I_k, \nu)$ can be calculated using the representation
\begin{equation}
\label{decomp_t_dist}
x \stackrel{d}{=} \sqrt{\tau}\, z,
\end{equation}
where
\[
z \sim N_k(0, I_k), \qquad 
\tau = \frac{\nu}{U}, \qquad 
U \sim \chi^2_\nu, \qquad \text{$z$ and $U$ are independent.}
\]
The moments of $\tau$ are given by
\begin{align*}
&E[\tau] = \frac{\nu}{\nu - 2}(\triangleq \lambda_1), \quad
E[\tau^2] = \frac{\nu^2}{(\nu - 2)(\nu - 4)}(\triangleq \lambda_2), \\
&E[\tau^3] = \frac{\nu^3}{(\nu - 2)(\nu - 4)(\nu - 6)}(\triangleq \lambda_3),
\end{align*}
which are well defined for $\nu > 6$. Using \eqref{Wick_formula} and \eqref{decomp_t_dist}, we can readily derive the moments of $x$ when $x \sim  t_k(0, I_k, \nu)$. For mutually distinct indices $1\leq i , j , l , m,  s, t \leq k$, the following equations hold:

\begin{equation}
\label{moments_t_dist}
\begin{aligned}
&E[x_i] = 0,\\
&E[x_i^2] = \lambda_1,\quad E[x_i x_j] = 0, \\
&E[x_i^3] = 0, \quad E[x_i^2 x_j] =0,\quad E[x_i x_j x_l] =0, \\
&E[x_i^4] = 3\lambda_2,\quad E[x_i^3 x_j] =0,\quad E[x_i^2 x_j^2]= \lambda_2,\quad E[x_i^2 x_j  x_l] =0,\quad E[x_i x_j x_l x_m]=0,\\
&E[x_i^5] = 0, \quad E[x_i^4 x_j] =0,\quad E[x_i^3 x_j^2]=0,\quad E[x_i^3 x_j x_l]=0,\quad E[x_i^2 x_j^2 x_l]=0,\\
&E[x_i^2 x_j x_l x_m] = 0,\quad E[x_i x_j x_l x_m x_s]=0,\\
&E[x_i^6] = 15\lambda_3, \quad E[x_i^5 x_j] =0,\quad E[x_i^4 x_j^2]= 3\lambda_3,\quad E[x_i^3 x_j^3]=0, \quad E[x_i^4 x_j x_l]=0,\\ &E[x_i^3 x_j^2 x_l]= 0,\quad E[x_i^2 x_j^2 x_l^2]=\lambda_3, \quad E[x_i^3 x_j x_l x_m] = 0, \quad E[x_i^2 x_j^2 x_l x_m]=0,\\
&E[x_i^2 x_j x_l x_m x_s]=0,\quad E[x_i x_j x_l x_m x_s x_t]=0.
\end{aligned}
\end{equation}
%
%
%
%
%
\section{Metrics in Theorem 2 of \cite{Sheena4} }
\label{metrics_Th3}
First we calculate $G^*(\theta_*)=\tilde{G}(\theta_*)$ and $G(\theta^*)$. Notice that $G^*(=\tilde{G})$ and $G$ are the variance-covariance matrices of $\xi(x)$ under the conditions respectively $x \sim N_k(0, \lambda_1 I_k)$ and $x \sim  t_k(0, I_k, \nu)$.
In order to avoid confusion, we redefine the notations; 
\begin{align*}
&\tilde{g}_{ij} \triangleq Cov(\xi_i, \xi_j) = Cov(x_i, x_j), \quad 1\leq i, j \leq k,\\
&\tilde{g}_{(i,j)l} = \tilde{g}_{l(i,j)} \triangleq Cov(\xi_{ij}, \xi_{l}) = Cov(x_i x_j, x_l), \quad 1\leq i \leq j \leq k,\ 1\leq l \leq k,\\
&\tilde{g}_{(i,j)(l,m)} \triangleq Cov(\xi_{ij}, \xi_{lm}) = Cov(x_i x_j, x_l x_m), \quad 1\leq i \leq j \leq k,\ 1\leq l \leq m \leq k,
\end{align*}
under the condition $x \sim N_k(0, \lambda_1 I_k)$, while
\begin{align*}
&{g}_{ij} \triangleq Cov(\xi_i, \xi_j)= Cov(x_i, x_j), \quad 1\leq i, j \leq k,\\
&{g}_{(ij)l} = g_{l(i,j)} \triangleq Cov(\xi_{ij}, \xi_{l}) = Cov(x_i x_j, x_l), \quad 1\leq i \leq j \leq k,\ 1\leq l \leq k,\\
&{g}_{(ij)(lm)} \triangleq Cov(\xi_{ij}, \xi_{lm}) = Cov(x_i x_j, x_l x_m)\quad 1\leq i \leq j \leq k,\ 1\leq l \leq m \leq k,
\end{align*}
under the condition $x \sim  t_k(0, I_k, \nu)$. Let 
\[
\tilde{G} \triangleq 
\left(
\begin{array}{c|c}
\bigl(\tilde{g}_{ij}\bigr)_{1\leq i,j \leq k} &\bigl(\tilde{g}_{l(ij)}\bigr)_{1\leq i \leq j \leq k, 1\leq l \leq k} \\
\hline
\bigl(\tilde{g}_{(ij)l}\bigr)_{1\leq i \leq j \leq k, 1\leq l \leq k} & \bigl(\tilde{g}_{(ij)(lm)}\bigr)_{1\leq i \leq j \leq k, 1\leq l \leq m \leq k} 
\end{array}
\right)
\]
\[
\tilde{G}^{-1} \triangleq 
\left(
\begin{array}{c|c}
\bigl(\tilde{g}^{ij}\bigr)_{1\leq i,j \leq k} &\bigl(\tilde{g}^{l(ij)}\bigr)_{1\leq i \leq j \leq k, 1\leq l \leq k} \\
\hline
\bigl(\tilde{g}^{(ij)l}\bigr)_{1\leq i \leq j \leq k, 1\leq l \leq k} & \bigl(\tilde{g}^{(ij)(lm)}\bigr)_{1\leq i \leq j \leq k, 1\leq l \leq m \leq k} 
\end{array}
\right)
\]
\[
G \triangleq 
\left(
\begin{array}{c|c}
\bigl(g_{ij}\bigr)_{1\leq i,j \leq k} &\bigl(g_{l(ij)}\bigr)_{1\leq i \leq j \leq k, 1\leq l \leq k} \\
\hline
\bigl(g_{(ij)l}\bigr)_{1\leq i \leq j \leq k, 1\leq l \leq k} & \bigl(g_{(ij)(lm)}\bigr)_{1\leq i \leq j \leq k, 1\leq l \leq m \leq k} 
\end{array}
\right)
\]
For $\tilde{G}$, we obtain the following results.
\begin{align*}
& \tilde{g}_{ij} = \lambda_1 \delta_{ij}, \quad i,j =1,\ldots,k, \quad \text{$\delta_{ij}$ is Kronecker delta,}\\
& \tilde{g}_{(ij)l} = E(x_i x_j x_l) - E(x_i x_j)E(x_l)= 0,\\
& \tilde{g}_{(i,j)(l,m)}  = E(x_i x_j x_l x_m) - E(x_i x_j)E(x_l x_m) \\
&=
\begin{cases}
2\lambda_1^2&\text{if $1\le i=j=l=m \leq k,$} \\
\lambda_1^2 &\text{if $1\leq i=l < j=m \leq k,$} \\
0 &\text{otherwise,}
\end{cases}
\end{align*}
The matrix $\tilde{G}$ is diagonal, and $\tilde{G}^{-1}$ is given as follows:
\begin{align*}
&\tilde{g}^{ij} = \lambda_1^{-1}\delta_{ij},\quad 1\leq i, j \leq k, \\
&\tilde{g}^{l(ij)}=\tilde{g}^{(ij)l}=0,\quad 1\leq l \leq k,\ 1\leq i \leq j \leq k, \\
&\tilde{g}^{(ij)(lm)}=(1+\delta_{ij})^{-1}\lambda_1^{-2} \delta_{il}\delta_{jm},\quad  1\leq i \leq j \leq k,\  1\leq l \leq m \leq k.
\end{align*}
The matrix $G(\theta_*)$ is calculated from \eqref{moments_t_dist}.
For $1\leq i \leq j \leq  k$, $1 \leq l \leq m \leq k$, we have
\begin{align*}
&Cov(\xi_i, \xi_j) = Cov(x_i, x_j) = E[x_i x_j] - E[x_i]E[x_j] =\lambda_1 \delta_{ij}, \\
&Cov(\xi_l, \xi_{ij}) = Cov(x_l, x_i x_j) = E[x_i x_j x_l] - E[x_l]E[x_i x_j]= 0, \\
&Cov(\xi_{ij}, \xi_{lm}) = Cov(x_i x_j, x_l x_m) = E[x_i x_j x_l x_m] - E[x_i x_j]E[x_l x_m] \\
&=
\begin{cases}
3\lambda_2-\lambda_1^2&\text{if  $i=j=l=m,$} \\
\lambda_2 &\text{if $i=l < j=m,$} \\
\lambda_2-\lambda_1^2&\text{if $i=j, l=m, i\ne l,$} \\
0 &\text{otherwise,}
\end{cases}
\end{align*}
Consequently, $G(\theta_*)$ is given as follows:
\begin{align*}
&g_{ij}= \lambda_1\delta_{ij},\quad 1\leq i, j \leq k, \\
&g_{l(ij)}=g_{(ij)l}=0,\quad 1\leq l \leq k,\ 1\leq i \leq j \leq k, \\
&g_{(ij)(lm)} =
\begin{cases}
3\lambda_2-\lambda_1^2 &\text{if $1\leq i=j=l=m \leq k,$} \\
\lambda_2 &\text{if $1 \leq i = l < j = m \leq k,$} \\
\lambda_2-\lambda_1^2 &\text{if $1 \leq i=j \ne l = m \leq k$}\\
0 &\text{otherwise}
\end{cases}
\end{align*}
\section{Cumulants in Theorem 2 of \cite{Sheena4}}
\label{cumu_Th3}
We first redefine the notation for the third-order cumulants:
\begin{align*}
&\text{For $ 1\leq i,j,l \leq k$,}\nonumber\\
&\kappa_{ijl} \triangleq E\bigl[(\xi_i-E[\xi_i])(\xi_j-E[\xi_j])(\xi_l-E[\xi_l])\bigr].\\
&\text{For $1\leq i \leq j \leq k,  1\leq l, m\leq k$,}\nonumber\\
&\kappa_{(i j)l m} \triangleq E\bigl[(\xi_{ij}-E[\xi_{ij}])(\xi_l-E[\xi_l])(\xi_m-E[\xi_m])\bigr].\\
&\text{For $1\leq i \leq j \leq k,  1\leq l \leq m \leq k, 1\leq s \leq k$,}\nonumber\\
&\kappa_{(i j)(l m)s} \triangleq E\bigl[(\xi_{ij}-E[\xi_{ij}])(\xi_{lm}-E[\xi_{lm}])(\xi_{s}-E[\xi_{s}])\bigr].\\
&\text{For $1\leq i \leq j \leq k,  1\leq l \leq m \leq k, 1\leq s \leq t \leq k$,}\nonumber\\
&\kappa_{(i j)(l m)(s t)} = E\bigl[(\xi_{ij}-E[\xi_{ij}])(\xi_{lm}-E[\xi_{lm}])(\xi_{st}-E[\xi_{st}])\bigr]
\end{align*}
where the expectation is taken under the distribution $x \sim  t_k(0, I_k, \nu)$. The quantities $\kappa_{ijl}^*$, $\kappa^*_{(i j)l m}$, $\kappa^*_{(i j)(l m)s}$, and $\kappa^*_{(i j)(l m)(s t)}$ are defined analogously under the distribution $x \sim  N_k(0, \lambda_1 I_k)$.

The notation for the fourth-order cumulants is defined as follows, where all expectations are taken under the distribution $x \sim  N_k(0, \lambda_1 I_k)$:
\begin{align*}
&\text{For $ 1\leq i,j,l,m \leq k$,}\nonumber\\
&\kappa^*_{ijlm}\\
& \triangleq E\bigl[(\xi_i-E[\xi_i])(\xi_j-E[\xi_j])(\xi_l-E[\xi_l])(\xi_m-E[\xi_m])\bigr]\\
&\qquad -E\bigl[(\xi_i-E[\xi_i])(\xi_j-E[\xi_j])\bigr]E\bigl[(\xi_l-E[\xi_l])(\xi_m-E[\xi_m])\bigr]\\
&\qquad -E\bigl[(\xi_i-E[\xi_i])(\xi_l-E[\xi_l])\bigr]E\bigl[(\xi_j-E[\xi_j])(\xi_m-E[\xi_m])\bigr]\\
&\qquad -E\bigl[(\xi_i-E[\xi_i])(\xi_m-E[\xi_m])\bigr]E\bigl[(\xi_j-E[\xi_j])(\xi_l-E[\xi_l])\bigr].\\
&\text{For $ 1\leq i \leq j \leq k, 1\leq l, m, s \leq k$,}\nonumber\\
&\kappa^*_{(i j)l m s} \\
&\triangleq E\bigl[(\xi_{ij}-E[\xi_{ij}])(\xi_l-E[\xi_l])(\xi_m-E[\xi_m])(\xi_s-E[\xi_s])\bigr].\\
&\qquad -E\bigl[(\xi_{ij}-E[\xi_{ij}])(\xi_l-E[\xi_l])\bigr]E\bigl[(\xi_m-E[\xi_m])(\xi_s-E[\xi_s])\bigr]\\
&\qquad -E\bigl[(\xi_{ij}-E[\xi_{ij}])(\xi_m-E[\xi_m])\bigr]E\bigl[(\xi_l-E[\xi_l])(\xi_s-E[\xi_s])\bigr]\\
&\qquad -E\bigl[(\xi_{ij}-E[\xi_{ij}])(\xi_s-E[\xi_s])\bigr]E\bigl[(\xi_l-E[\xi_l])(\xi_m-E[\xi_m])\bigr].\\
&\text{For $ 1\leq i \leq j \leq k, 1\leq l \leq m \leq k,  1\leq s, t \leq k$,}\nonumber\\
&\kappa^*_{(i j)(lm)s t} \\
&\triangleq E\bigl[(\xi_{ij}-E[\xi_{ij}])(\xi_{lm}-E[\xi_{lm}])(\xi_s-E[\xi_s])(\xi_t-E[\xi_t])\bigr]\\
&\qquad -E\bigl[(\xi_{ij}-E[\xi_{ij}])(\xi_{lm}-E[\xi_{lm}])\bigr]E\bigl[(\xi_s-E[\xi_s])(\xi_t-E[\xi_t])\bigr]\\
&\qquad -E\bigl[(\xi_{ij}-E[\xi_{ij}])(\xi_s-E[\xi_s])\bigr]E\bigl[(\xi_{lm}-E[\xi_{lm}])(\xi_t-E[\xi_t])\bigr]\\
&\qquad -E\bigl[(\xi_{ij}-E[\xi_{ij}])(\xi_t-E[\xi_t])\bigr]E\bigl[(\xi_{lm}-E[\xi_{lm}])(\xi_s-E[\xi_s])\bigr].\\
&\text{For $1\leq i \leq j \leq k,  1\leq l \leq m \leq k, 1\leq s \leq t \leq k, 1\leq u \leq k$,}\nonumber\\
&\kappa^*_{(i j)(l m)(s t) u}\\
&\triangleq E\bigl[(\xi_{ij}-E[\xi_{ij}])(\xi_{lm}-E[\xi_{lm}])(\xi_{st}-E[\xi_{st}])(\xi_u-E[\xi_u])\bigr]\\
&\qquad -E\bigl[(\xi_{ij}-E[\xi_{ij}])(\xi_{lm}-E[\xi_{lm}])\bigr]E\bigl[(\xi_{st}-E[\xi_{st}])(\xi_u-E[\xi_u])\bigr]\\
&\qquad -E\bigl[(\xi_{ij}-E[\xi_{ij}])(\xi_{st}-E[\xi_{st}])\bigr]E\bigl[(\xi_{lm}-E[\xi_{lm}])(\xi_u-E[\xi_u])\bigr]\\
&\qquad -E\bigl[(\xi_{ij}-E[\xi_{ij}])(\xi_u-E[\xi_u])\bigr]E\bigl[(\xi_{lm}-E[\xi_{lm}])(\xi_{st}-E[\xi_{st}])\bigr].\\
&\text{For $1\leq i \leq j \leq k,  1\leq l \leq m \leq k, 1\leq s \leq t \leq k, 1\leq u \leq v \leq k$,}\nonumber\\
&\kappa^*_{(i j)(l m)(s t)(u v)}\\
&\triangleq E\bigl[(\xi_{ij}-E[\xi_{ij}])(\xi_{lm}-E[\xi_{lm}])(\xi_{st}-E[\xi_{st}])(\xi_{uv}-E[\xi_{uv}])\bigr]\\
&\qquad -E\bigl[(\xi_{ij}-E[\xi_{ij}])(\xi_{lm}-E[\xi_{lm}])\bigr]E\bigl[(\xi_{st}-E[\xi_{st}])(\xi_{uv}-E[\xi_{uv}])\bigr]\\
&\qquad -E\bigl[(\xi_{ij}-E[\xi_{ij}])(\xi_{st}-E[\xi_{st}])\bigr]E\bigl[(\xi_{lm}-E[\xi_{lm}])(\xi_{uv}-E[\xi_{uv}])\bigr]\\
&\qquad -E\bigl[(\xi_{ij}-E[\xi_{ij}])(\xi_{uv}-E[\xi_{uv}])\bigr]E\bigl[(\xi_{lm}-E[\xi_{lm}])(\xi_{st}-E[\xi_{st}])\bigr].\\
\end{align*}

\subsection{Calculation of third-order cumulants}
\mbox{}\par
The first three third-order cumulants, $\kappa_{ijl}$, $\kappa_{(ij)lm}$, and $\kappa_{(ij)(lm)s}$, are readily obtained as follows:
\begin{align*}
&\kappa_{ijl} = E[x_i x_j x_l] =0,\\
&\kappa_{(ij)lm} = E[(x_i x_j - \delta_{ij}\lambda_1)x_l x_m] = E[x_i x_j x_l x_m ] - \delta_{ij} \lambda_1E[x_l x_m] \\
&= 
\begin{cases}
3\lambda_2-\lambda_1^2, \text{ if $i=j=l=m$,}\\
\lambda_2 - \lambda_1^2,\text{ if $i=j\ne l=m$,}\\
\lambda_2, \text{ if $i=l < j=m$,}\\
0,\text{otherwise},
\end{cases}
\\
&\kappa_{(ij)(lm)s}=E\bigl[(\xi_{ij}-E[\xi_{ij}])(\xi_{lm}-E[\xi_{lm}])\xi_s\bigr]\\
&= E\bigl[(x_i x_j - \delta_{ij}\lambda_1)(x_l x_m - \delta_{lm}\lambda_1)x_s\bigr]\\
&= E[x_i x_j x_l x_m x_s]
- \delta_{lm}\lambda_1 E[x_i x_j x_s]
- \delta_{ij}\lambda_1 E[x_l x_m x_s]
+ \delta_{ij}\delta_{lm}\lambda_1^2 E[x_s]= 0
\end{align*}

We now focus on the cumulant $\kappa_{(ij)(lm)(st)}$.
\begin{align}
&\kappa_{(ij)(lm)(st)}=E\bigl[(\xi_{ij}-E[\xi_{ij}])(\xi_{lm}-E[\xi_{lm}])(\xi_{st}-E[\xi_{st}])\bigr]\nonumber\\
&= E\bigl[(x_i x_j - \lambda_1 \delta_{ij})(x_l x_m - \lambda_1 \delta_{lm})(x_s x_t - \lambda_1 \delta_{st})\bigr]\nonumber\\
&= E[x_i x_j x_l x_m x_s x_t]
- \lambda_1 \delta_{ij} E[x_l x_m x_s x_t]
- \lambda_1 \delta_{lm} E[x_i x_j x_s x_t]
- \lambda_1 \delta_{st} E[x_i x_j x_l x_m]\nonumber\\
&\qquad+ \lambda_1^2 \delta_{ij}\delta_{lm} E[x_s x_t]
+ \lambda_1^2 \delta_{ij}\delta_{st} E[x_l x_m]
+ \lambda_1^2 \delta_{lm}\delta_{st} E[x_i x_j]
- \lambda_1^3 \delta_{ij}\delta_{lm}\delta_{st} \nonumber\\
&= E[x_i x_j x_l x_m x_s x_t]
- \lambda_1 \delta_{ij} E[x_l x_m x_s x_t]
- \lambda_1 \delta_{lm} E[x_i x_j x_s x_t]
- \lambda_1 \delta_{st} E[x_i x_j x_l x_m]\nonumber\\
&\qquad+2 \lambda_1^3 \delta_{ij}\delta_{lm}\delta_{st}. \label{3rd_cumulant_(ij)(lm)(st)}
\end{align}
We classify all index patterns $(i,j,l,m,s,t)$ subject to the constraints
\[
1\leq i \le j \leq k,\qquad 1 \leq l \le m \leq k, \qquad 1\leq s \le t \leq k.
\]
We group the patterns according to the number of distinct values among
\[
(i,j,l,m,s,t),
\]
and by
\[
r = \delta_{ij} + \delta_{lm} + \delta_{st}.
\]
\\
{\bf Class A (one distinct value)}\\
All six indices are identical: 
\[
i = j = l = m = s = t.
\]
\bigskip
\\
{\bf Class B1 (two distinct values, $r=3$)}\\
The admissible patterns are:
\[
\begin{aligned}
&i = j = l = m,\quad s = t,\quad i \neq s,\\
&i = j = s = t,\quad l = m,\quad i \neq l,\\
&l = m = s = t,\quad i = j,\quad l \neq i.
\end{aligned}
\]
{\bf Class B2 (two distinct values, $r\ne 3$)}\\
The admissible patterns are:
\[
\begin{aligned}
&i = j = l = s,\quad m = t,\quad i \neq m,\\
&l = s,\quad i = j = m = t,\quad l \neq i,\\
&i = l = m = s,\quad j = t,\quad i \neq j,\\
&i = l = s = t,\quad j = m,\quad i \neq j,\\
&i = l,\quad j = m = s = t,\quad i \neq j,\\
&i = s,\quad j = l = m = t,\quad i \neq j.
\end{aligned}
\]
{\bf Class C(3) (three distinct values, $r=3$)}\\
The only pattern is:
\[
i = j,\quad l = m,\quad s = t,
\]
with the three values mutually distinct.
\\
\\
{\bf Class C(2) (three distinct values, $r=2$)}\\
This class is empty.
\\
\\
{\bf Class C(1) (three distinct values, $r=1$)}\\
The admissible patterns are:
\[
\begin{aligned}
&i = j,\quad l = s,\quad m = t,\\
&i = l,\quad j = m,\quad s = t,\\
&i = s,\quad j = t,\quad l = m,
\end{aligned}
\]
where the three values are mutually distinct.
\\
\\
{\bf Class C(0) (three distinct values, $r=0$)}\\
The admissible patterns are:
\[
\begin{aligned}
&i = l,\quad j = s,\quad m = t,\\
&i = l,\quad j = t,\quad m = s,\\
&i = s,\quad j = l,\quad m = t,\\
&i = m,\quad j = t,\quad l = s,\\
&i = s,\quad j = m,\quad l = t,\\
&i = t,\quad j = m,\quad l = s,
\end{aligned}
\]
where the three values are mutually distinct.

For each class, substituting the moments of $x$ into \eqref{3rd_cumulant_(ij)(lm)(st)} yields the values of $\kappa_{(ij)(lm)(st)}$ shown in Table \ref{table_kappa_(ij)(lm)(st)}.

\renewcommand{\arraystretch}{1.5}
\begin{table}[htbp]
\caption{Values and counts of $\kappa_{(ij)(lm)(st)}$}
\label{table_kappa_(ij)(lm)(st)}
\centering
\begin{tabular}{|l|l|l|l|}
\hline
Class & Value of $\kappa_{(i,j)(l,m)(s,t)}$ & Number of Index Patterns  \\
\hline
A &
$15\lambda_3 - 9\lambda_1 \lambda_2 + 2\lambda_1^3$ &
$k$ \\
\hline
B1 &
$3\lambda_3 - 5\lambda_1 \lambda_2 + 2\lambda_1^3$ &
$3k(k-1)$ \\
\hline
B2 &
$3\lambda_3 - \lambda_1 \lambda_2$ &
$3k(k-1)$ \\
\hline
C(3) &
$\lambda_3 - 3\lambda_1 \lambda_2 + 2\lambda_1^3$ &
$k(k-1)(k-2)$ \\
\hline
C(1) &
$\lambda_3 - \lambda_1 \lambda_2$ &
$\dfrac{3}{2}k(k-1)(k-2)$ \\
\hline
C(0) &
$\lambda_3$ &
$k(k-1)(k-2)$ \\
\hline
Others &
$0$ & \text{unknown}
\\
\hline
\end{tabular}
\end{table}

Next, we calculate $\kappa_{ijl}^*$, $\kappa^*_{(i j)l m}$, $\kappa^*_{(i j)(l m)s}$, and $\kappa^*_{(i j)(l m)(s t)}$. Because $x \sim  N_k(0, \lambda_1I_k)$ can be represented as
\begin{equation}
\label{decomp_x_dist}
x \stackrel{d}{=} \sqrt{\tau}\, z,
\end{equation}
where
\[
z \sim N_k(0, I_k), \qquad \tau = \lambda_1, \qquad 
\]
it suffices to modify the results for $\kappa_{ijl}$, $\kappa_{(i j)l m}$, $\kappa_{(i j)(l m)s}$, and $\kappa_{(i j)(l m)(s t)}$ by making the substitutions
\[
\lambda_2 \rightarrow \lambda_1^2,\qquad \lambda_3 \rightarrow \lambda_1^3.
\]
Consequently, we obtain the following results:
\begin{align*}
&\kappa^*_{ijl} =0,\\
&\kappa^*_{(ij)lm}= 
\begin{cases}
2\lambda_1^2, \text{ if $i=j=l=m$,}\\
\lambda_1^2, \text{ if $i=l < j=m$,}\\
0,\text{otherwise},
\end{cases}
\\
&\kappa_{(ij)(lm)s}=0.
\end{align*}
$\kappa^*_{(ij)(lm)(st)}$ is given in Table \ref{table*(ij)(lm)(st)}.
\begin{table}[htbp]
\caption{Values and  counts of $\kappa^*_{(ij)(lm)(st)}$}
\label{table*(ij)(lm)(st)}
\centering
\begin{tabular}{|l|l|l|l|}
\hline
Class & Value of $\kappa^*_{(i,j)(l,m)(s,t)}$ & Number of Index Patterns  \\
\hline
A &
$8\lambda_1^3$ &
$k$ \\
\hline
B2 &
$2\lambda_1^3$ &
$3k(k-1)$ \\
\hline
C(0) &
$\lambda_1^3$ &
$k(k-1)(k-2)$ \\
\hline
Others &
$0$ & \text{unknown}
\\
\hline
\end{tabular}
\end{table}

\subsection{Calculation of $\kappa^*_{ijlm}$, $\kappa^*_{(i j)l m s} $, $\kappa^*_{(i j)(l m)(s t) u}$}
\mbox{}\par
We calculate the fourth-order cumulants $\kappa^*_{ijlm}$, $\kappa^*_{(i j)l m s} $, and $\kappa^*_{(i j)(l m)(s t) u}$.

It follows immediately from \eqref{Wick_formula} that these cumulants vanish.
\begin{align*}
&\text{For $ 1\leq i,j,l,m \leq k$,}\\
&\kappa^*_{ijlm}\\
&=E[x_i x_j x_l x_m]-E[x_i x_j]E[x_l x_m]-E[x_i x_l]E[x_j x_m]-E[x_i x_m]E[x_j x_l]\\
&=\lambda_1^2(\delta_{ij}\delta_{lm}+\delta_{il}\delta_{jm}+\delta_{im}\delta_{jl})-\lambda_1^2\delta_{ij}\delta_{lm}-\lambda_1^2\delta_{il}\delta_{jm}-\lambda_1^2\delta_{im}\delta_{jl}\\
&=0.\\
&\text{For $ 1\leq i \leq j \leq k, 1\leq l, m, s \leq k$,}\\
&\kappa^*_{(i j)l m s} \\
&= E\bigl[(x_i x_j-E[x_i x_j])(x_l-E[x_l])(x_m-E[x_m])(x_s-E[x_s])\bigr].\\
&\qquad -E\bigl[(x_i x_j-E[x_i x_j])(x_l-E[x_l])\bigr]E\bigl[(x_m-E[x_m])(x_s-E[x_s])\bigr]\\
&\qquad -E\bigl[(x_i x_j-E[x_i x_j])(x_m-E[x_m])\bigr]E\bigl[(x_l-E[x_l])(x_s-E[x_s])\bigr]\\
&\qquad -E\bigl[(x_i x_j-E[x_i x_j])(x_s-E[x_s])\bigr]E\bigl[(x_l-E[x_l])(x_m-E[x_m])\bigr]\\
&= E\bigl[(x_i x_j-\lambda_1\delta_{ij})x_l x_m x_s\bigr].\\
&\qquad -E\bigl[(x_i x_j-\lambda_1\delta_{ij})x_l])\bigr]E\bigl[x_m x_s\bigr]\\
&\qquad -E\bigl[(x_i x_j-\lambda_1\delta_{ij})x_m\bigr]E\bigl[x_l x_s\bigr]\\
&\qquad -E\bigl[(x_i x_j-\lambda_1\delta_{ij})x_s\bigr]E\bigl[x_l x_m\bigr]\\
&=0.\\
&\text{For $1\leq i \leq j \leq k,  1\leq l \leq m \leq k, 1\leq s \leq t \leq k, 1\leq u \leq k$,}\nonumber\\
&\kappa^*_{(i j)(l m)(s t) u}\\
&\triangleq E\bigl[(x_i x_j-E[x_i x_j])(x_l x_m-E[x_l x_m])(x_s x_t-E[x_s x_t])(x_u-E[x_u])\bigr]\\
&\qquad -E\bigl[(x_i x_j-E[x_i x_j])(x_l x_m-E[x_l x_m])\bigr]E\bigl[(x_s x_t-E[x_s x_t])(x_u-E[x_u])\bigr]\\
&\qquad -E\bigl[(x_i x_j-E[x_i x_j])(x_s x_t-E[x_s x_t])\bigr]E\bigl[(x_l x_m-E[x_l x_m])(x_u-E[x_u])\bigr]\\
&\qquad -E\bigl[(x_i x_j-E[x_i x_j])(x_u-E[x_u])\bigr]E\bigl[(x_l x_m-E[x_l x_m])(x_s x_t-E[x_s x_t])\bigr].\\
&= E\bigl[(x_i x_j-\lambda_1\delta_{ij})(x_l x_m-\lambda_1\delta_{lm})(x_s x_t-\lambda_1\delta_{st})x_u\bigr]\\
&\qquad -E\bigl[(x_i x_j-\lambda_1\delta_{ij})(x_l x_m-\lambda_1\delta_{lm})\bigr]E\bigl[(x_s x_t-\lambda_1\delta_{st})x_u\bigr]\\
&\qquad -E\bigl[(x_i x_j-\lambda_1\delta_{ij})(x_s x_t-\lambda_1\delta_{st})\bigr]E\bigl[(x_l x_m-\lambda_1\delta_{lm})x_u\bigr]\\
&\qquad -E\bigl[(x_i x_j-\lambda_1\delta_{ij})x_u\bigr]E\bigl[(x_l x_m-\lambda_1\delta_{lm})(x_s x_t-\lambda_1\delta_{st})\bigr]\\
&=0.
\end{align*}

\subsection{Calculation of $\kappa^*_{(i j)(lm)s t}$}
\mbox{}\par
We now calculate $\kappa^*_{(i j)(lm)s t}$ for $ 1\leq i \leq j \leq k, 1\leq l \leq m \leq k,  1\leq s, t \leq k$.
\begin{align*}
&\kappa^*_{(i j)(lm)s t} \\
&\triangleq E\bigl[(x_i x_j-\lambda_1 \delta_{ij})(x_l x_m-\lambda_1 \delta_{lm})x_s x_t\bigr]\\
&\qquad -E\bigl[(x_i x_j-\lambda_1 \delta_{ij})(x_l x_m-\lambda_1 \delta_{lm})\bigr]E\bigl[x_s x_t\bigr]\\
&\qquad -E\bigl[(x_i x_j-\lambda_1 \delta_{ij})x_s\bigr]E\bigl[(x_l x_m-\lambda_1 \delta_{lm})x_t\bigr]\\
&\qquad -E\bigl[(x_i x_j-\lambda_1 \delta_{ij})x_t\bigr]E\bigl[(x_l x_m-\lambda_1 \delta_{lm})x_s\bigr].\\
&= E\bigl[(x_i x_j-\lambda_1 \delta_{ij})(x_l x_m-\lambda_1 \delta_{lm})x_s x_t\bigr]\\
&\qquad -E\bigl[(x_i x_j-\lambda_1 \delta_{ij})(x_l x_m-\lambda_1 \delta_{lm})\bigr]E[x_s x_t]
\end{align*}
We expand the terms as follows:
\[
\begin{aligned}
&E\bigl[ (x_i x_j - \lambda_1 \delta_{ij})(x_l x_m - \lambda_1 \delta_{lm}) x_s x_t\bigr]\\
&=E[x_i x_j x_l x_m x_s x_t]\\
&\quad- \lambda_1 \delta_{lm} E[x_i x_j x_s x_t]\\
&\quad- \lambda_1 \delta_{ij} E[x_l x_m x_s x_t]\\
&\quad+ \lambda_1^2 \delta_{ij}\delta_{lm} E[x_s x_t]
\end{aligned}
\]
\begin{align*}
&E[(x_i x_j - \lambda_1 \delta_{ij})(x_l x_m - \lambda_1 \delta_{lm})]E[x_s x_t]\\
&=\lambda_1 \delta_{st}E[x_i x_j x_l x_m] - \lambda_1^3 \delta_{ij}\delta_{lm}\delta_{st}.
\end{align*}
Thus,
\begin{equation}
\label{simple_form_k*(ij)(lm)st}
\begin{aligned}
\kappa^*_{(ij)(lm)st}
={}&
E[x_i x_j x_l x_m x_s x_t]\\
&- \lambda_1 \delta_{lm} E[x_i x_j x_s x_t]\\
&- \lambda_1 \delta_{ij} E[x_l x_m x_s x_t]\\
&- \lambda_1 \delta_{st} E[x_i x_j x_l x_m]\\
&+ 2 \lambda_1^3 \delta_{ij}\delta_{lm}\delta_{st}.
\end{aligned}
\end{equation}

By \eqref{Wick_formula}, the fourth- and sixth-order moments of $x$ can be expressed using $\delta$ notation as follows:
\begin{equation}
\label{forth-order_normal}
E[x_i x_j x_l x_m] = \lambda_1^2 \, S_4(i,j,l,m),
\end{equation}
where $S_4$ is the sum over all pairings of four indices, i.e.
\[
S_4(i, j, l, m)
= \delta_{ij}\delta_{lm}
+ \delta_{il}\delta_{jm}
+ \delta_{im}\delta_{jl}.
\]
\begin{equation}
\label{sixth-order_normal}
E[x_i x_j x_l x_m x_s x_t]
= \lambda_1^3 \, S_6(i,j,l,m,s,t),
\end{equation}
where $S_6$ is the sum over all pairings of six indices, i.e.
\[
\begin{aligned}
S_6(i,j,l,m,s,t) =\;&
\delta_{ij}\delta_{lm}\delta_{st}
+\delta_{ij}\delta_{ls}\delta_{mt}
+\delta_{ij}\delta_{lt}\delta_{ms}\\
&+\delta_{il}\delta_{jm}\delta_{st}
+\delta_{il}\delta_{js}\delta_{mt}
+\delta_{il}\delta_{jt}\delta_{ms}\\
&+\delta_{im}\delta_{jl}\delta_{st}
+\delta_{im}\delta_{js}\delta_{lt}
+\delta_{im}\delta_{jt}\delta_{ls}\\
&+\delta_{is}\delta_{jl}\delta_{mt}
+\delta_{is}\delta_{jm}\delta_{lt}
+\delta_{is}\delta_{jt}\delta_{lm}\\
&+\delta_{it}\delta_{jl}\delta_{ms}
+\delta_{it}\delta_{jm}\delta_{ls}
+\delta_{it}\delta_{js}\delta_{lm}.
\end{aligned}
\]

Substituting \eqref{forth-order_normal} and \eqref{sixth-order_normal} into \eqref{simple_form_k*(ij)(lm)st}, we observe the following cancellations. The coefficient of $\delta_{ij}\delta_{lm}\delta_{st}$ is
\[
1-1-1-1+2=0,
\]
and therefore vanishes. Similarly, the following six terms cancel pairwise:
\[
\begin{aligned}
&\delta_{ij}\delta_{ls}\delta_{mt},\quad
\delta_{ij}\delta_{lt}\delta_{ms},\\
&\delta_{is}\delta_{jt}\delta_{lm},\quad
\delta_{it}\delta_{js}\delta_{lm},\\
&\delta_{il}\delta_{jm}\delta_{st},\quad
\delta_{im}\delta_{jl}\delta_{st}.
\end{aligned}
\]

Hence only eight terms remain:
\begin{equation}
\label{new_def_kappa_(ij)(lm)st}
\begin{aligned}
\kappa^*(\triangleq \kappa^*_{(ij)(lm)st})
=
\lambda_1^3 (&
\delta_{il}\delta_{js}\delta_{mt}
+\delta_{il}\delta_{jt}\delta_{ms}
+\delta_{im}\delta_{js}\delta_{lt}
+\delta_{im}\delta_{jt}\delta_{ls}\\
&+\delta_{is}\delta_{jl}\delta_{mt}
+\delta_{it}\delta_{jl}\delta_{ms}
+\delta_{is}\delta_{jm}\delta_{lt}
+\delta_{it}\delta_{jm}\delta_{ls}
).
\end{aligned}
\end{equation}
Each term in \eqref{new_def_kappa_(ij)(lm)st} corresponds to three pairs. For example,
\[
\delta_{il}\delta_{js}\delta_{mt} \longleftrightarrow (i, l)(j, s)(m,t).
\]
Among all pairings of $(i, j, l, m, s, t)$, those appearing in \eqref{new_def_kappa_(ij)(lm)st} are characterized by the absence of the pairs $(i, j)$, $(l, m)$, and $(s,t)$.

We classify the index patterns $(i, j, l, m, s, t)$ according to the number of distinct indices and give the value of $\kappa^*$ for each class.\\
{\bf Class A (one distinct value)}\\
\[
i=j=l=m=s=t,\qquad \kappa^*=8\lambda_1^3.
\]
\\
{\bf Class B (two distinct values)}\\
Exactly two distinct values appear, yielding
\[
\kappa^*=2\lambda_1^3.
\]
This class contains the following 10 patterns. For $a < b$,
\begin{align*}
(i, j, l, m, s, t) =& (a, b, a, b, a, a), (a, b, a, b, b, b), \\
&(a, b, a, a, a, b), (a, b, a, a, b, a), (a, b, b, b, a, b), (a, b, b, b, b, a)\\
&(a, a, a, b, a, b), (a, a, a, b, b, a), (b, b, a, b, a, b), (b, b, a, b, b, a)
\end{align*}
\\
{\bf Class C (three distinct values)}\\
Exactly three distinct values appear (each twice), yielding
\[
\kappa^*=\lambda_1^3.
\]
This class contains the following 12 patterns. For $a < b <c$,
\begin{align*}
(i, j, l, m, s, t) =& (a, b, b, c, a, c), (a, b, b, c, c, a), (a, b, a, c, b, c), (a, b, a, c, c, b),\\
& (a, c, a, b, b, c), (a, c, a, b, c, b),(a, c, b, c, a, b), (a, c, b, c, b, a),\\
&(b, c, a, b, a, c), (b, c, a, b, c, a), (b, c, a, c, a, b), (b, c, a, c, b, a).
\end{align*}

For each class, the value of $\kappa^*_{(ij)(lm)st}$ and the number of patterns are summarized in Table \ref{table_kappa_(ij)(lm)st}.
%
%
%
%
%
\begin{table}[htbp]
\centering
\caption{Values and counts of $\kappa^*_{(ij)(lm)st}$}
\label{table_kappa_(ij)(lm)st}
\begin{tabular}{|c|c|c|}
\hline
Class & Value of  $\kappa_{(ij)(lm)st}$ & Number of Index Patterns \\
\hline
Class A & $8\lambda_1^3$ & $k$ \\
\hline
Class B & $2\lambda_1^3$ & $5k(k-1)$ \\
\hline
Class C & $\lambda_1^3$ & $2k(k-1)(k-2)$ \\
\hline
Others & $0$ & \text{unknown} \\
\hline
\end{tabular}
\end{table}

\subsection{Calculation of $\kappa^*_{(ij)(lm)(st)(uv)}$}
\mbox{}\par
Let
\begin{align*}
&A_{ij}=x_i x_j-\lambda_1\delta_{ij},
\qquad
A_{lm}=x_l x_m-\lambda_1\delta_{lm},
\\
&A_{st}=x_s x_t-\lambda_1\delta_{st},
\qquad
A_{uv}=x_u x_v-\lambda_1\delta_{uv}.
\end{align*}
Using \eqref{Wick_formula}, we can express
\[
\begin{aligned}
\kappa^*_{(ij)(lm)(st)(uv)}
&=
E[A_{ij}A_{lm}A_{st}A_{uv}]\\
&\quad-E[A_{ij}A_{lm}]E[A_{st}A_{uv}]\\
&\quad-E[A_{ij}A_{st}]E[A_{lm}A_{uv}]\\
&\quad-E[A_{ij}A_{uv}]E[A_{lm}A_{st}].
\end{aligned}
\]
as a sum of products of four $\delta$ functions.
We partition the eight indices into four \emph{blocks}
\[
B_1=\{i,j\},\quad B_2=\{l,m\},\quad B_3=\{s,t\},\quad B_4=\{u,v\}.
\]
A pairing of $\{i,j,l,m,s,t,u,v\}$ corresponds to a product of four Kronecker deltas; for example,
\[
(i,l)(j,s)(m,u)(t,v)
\;\longleftrightarrow\;
\delta_{il}\,\delta_{js}\,\delta_{mu}\,\delta_{tv}.
\]
Given a pairing $p$, construct a graph $G(p)$ whose vertices are the blocks
$\{B_1,B_2,B_3,B_4\}$, with an edge between two blocks whenever $p$
contains a pair connecting one index from each block.

\begin{itemize}
\item $p$ is called \emph{connected} if $G(p)$ is connected.
\item $p$ is called \emph{disconnected} if $G(p)$ is not connected.
\end{itemize}

In this cumulant-type expression, all disconnected pairings cancel after the products of lower-order moments are subtracted. We now illustrate this cancellation mechanism.

First, consider the cancellation that occurs when an internal pair is present. Suppose that a pairing contains the internal pair $(i,j)$.
\[
E[A_{ij}A_{lm}A_{st}A_{uv}] = E[x_i x_j A_{lm}A_{st}A_{uv}] - \lambda_1\delta_{ij}E[A_{lm}A_{st}A_{uv}]
\]
Choose a term from the expansion of $A_{lm}A_{st}A_{uv}$, say $C x_{a_1}\cdots x_{a_m}$, where $C$ is a constant. Consider
\[
C E[x_i x_j  x_{a_1} \cdots x_{a_m}] - C \lambda_1 \delta_{ij} E[x_{a_1} \cdots x_{a_m}] .
\]
Any term involving $\delta_{ij}$ in the first expression also appears in the second and therefore cancels. The same argument applies to the internal pairings
\[
(l,m), \qquad (s,t), \qquad (u,v).
\]
Hence, every pairing containing an internal pair vanishes from $E[A_{ij}A_{lm}A_{st}A_{uv}]$.
Similarly, $E[A_{ij}A_{lm}]$ contains no term involving an internal pair.
Consequently, $\kappa^*_{(ij)(lm)(st)(uv)}$ contains no internal pair in its $\delta$ representation.

Apart from pairings containing an internal pair, the only possible disconnected pairings are those that connect one pair of vertices among $B_1, B_2, B_3, B_4$ exclusively and connect the remaining pair exclusively. For example, consider the pairing
\[
(i,l)(j,m)(s,u)(t,v).
\]
This pairing connects $B_1$ only to $B_2$ and $B_3$ only to $B_4$.
It therefore decomposes into two independent components:
\[
\{B_1,B_2\}\quad\text{and}\quad \{B_3,B_4\}.
\]
Its contribution to the first expectation is
\[
\lambda_1^4\delta_{il}\delta_{jm}\delta_{su}\delta_{tv}.
\]

However, the same term also appears in
\[
E[A_{ij}A_{lm}]E[A_{st}A_{uv}].
\]
Indeed,
\[
E[A_{ij}A_{lm}]
=
\lambda_1^2(\delta_{il}\delta_{jm}+\delta_{im}\delta_{jl}),
\]
and
\[
E[A_{st}A_{uv}]
=
\lambda_1^2(\delta_{su}\delta_{tv}+\delta_{sv}\delta_{tu}).
\]
Hence, their product contains
\[
\lambda_1^4\delta_{il}\delta_{jm}\delta_{su}\delta_{tv}.
\]
Therefore, this contribution appears in $\kappa^*$ with coefficient
\[
1-1=0.
\]
Thus the disconnected pairing
\[
(i,l)(j,m)(s,u)(t,v)
\]
is cancelled.

Similarly, disconnected pairings of the type
\[
\{B_1,B_3\}\quad\text{and}\quad\{B_2,B_4\}
\]
are cancelled by
\[
E[A_{ij}A_{st}]E[A_{lm}A_{uv}],
\]
and disconnected pairings of the type
\[
\{B_1,B_4\}\quad\text{and}\quad\{B_2,B_3\}
\]
are cancelled by
\[
E[A_{ij}A_{uv}]E[A_{lm}A_{st}].
\]

Thus, all disconnected pairings cancel in the $\delta$ representation of $\kappa^*_{(ij)(lm)(st)(uv)}$. The admissible pairings are precisely those in which each block is connected to two distinct blocks, thereby forming a 4-cycle on $\{B_1,B_2,B_3,B_4\}$.
There are $3$ distinct cycles on four labeled blocks, and each cycle admits
$2^4=16$ choices, corresponding to the two ways of assigning the two indices in each block to its
incident edges. Hence, there are $3\times 16=48$ pairings in total. The complete list of 48 pairings is as follows:

\medskip
\noindent
\textbf{Cycle $B_1\!\!-\!B_2\!\!-\!B_4\!\!-\!B_3\!\!-\!B_1$:}
\[
\begin{aligned}
&(i,l)(j,s)(m,u)(t,v),\quad (i,l)(j,s)(m,v)(t,u),\\
&(i,l)(j,t)(m,u)(s,v),\quad (i,l)(j,t)(m,v)(s,u),\\
&(j,l)(i,s)(m,u)(t,v),\quad (j,l)(i,s)(m,v)(t,u),\\
&(j,l)(i,t)(m,u)(s,v),\quad (j,l)(i,t)(m,v)(s,u),\\
&(i,m)(j,s)(l,u)(t,v),\quad (i,m)(j,s)(l,v)(t,u),\\
&(i,m)(j,t)(l,u)(s,v),\quad (i,m)(j,t)(l,v)(s,u),\\
&(j,m)(i,s)(l,u)(t,v),\quad (j,m)(i,s)(l,v)(t,u),\\
&(j,m)(i,t)(l,u)(s,v),\quad (j,m)(i,t)(l,v)(s,u).
\end{aligned}
\]

\medskip
\noindent
\textbf{Cycle $B_1\!\!-\!B_2\!\!-\!B_3\!\!-\!B_4\!\!-\!B_1$:}
\[
\begin{aligned}
&(i,l)(j,u)(m,s)(v,t),\quad (i,l)(j,u)(m,t)(v,s),\\
&(i,l)(j,v)(m,s)(u,t),\quad (i,l)(j,v)(m,t)(u,s),\\
&(j,l)(i,u)(m,s)(v,t),\quad (j,l)(i,u)(m,t)(v,s),\\
&(j,l)(i,v)(m,s)(u,t),\quad (j,l)(i,v)(m,t)(u,s),\\
&(i,m)(j,u)(l,s)(v,t),\quad (i,m)(j,u)(l,t)(v,s),\\
&(i,m)(j,v)(l,s)(u,t),\quad (i,m)(j,v)(l,t)(u,s),\\
&(j,m)(i,u)(l,s)(v,t),\quad (j,m)(i,u)(l,t)(v,s),\\
&(j,m)(i,v)(l,s)(u,t),\quad (j,m)(i,v)(l,t)(u,s).
\end{aligned}
\]

\medskip
\noindent
\textbf{Cycle $B_1\!\!-\!B_3\!\!-\!B_2\!\!-\!B_4\!\!-\!B_1$:}
\[
\begin{aligned}
&(i,s)(j,u)(t,l)(m,v),\quad (i,s)(j,u)(t,m)(l,v),\\
&(i,s)(j,v)(t,l)(m,u),\quad (i,s)(j,v)(t,m)(l,u),\\
&(j,s)(i,u)(t,l)(m,v),\quad (j,s)(i,u)(t,m)(l,v),\\
&(j,s)(i,v)(t,l)(m,u),\quad (j,s)(i,v)(t,m)(l,u),\\
&(i,s)(j,u)(t,m)(l,v),\quad (i,s)(j,u)(t,l)(m,v),\\
&(i,s)(j,v)(t,m)(l,u),\quad (i,s)(j,v)(t,l)(m,u),\\
&(j,s)(i,u)(t,m)(l,v),\quad (j,s)(i,u)(t,l)(m,v),\\
&(j,s)(i,v)(t,m)(l,u),\quad (j,s)(i,v)(t,l)(m,u).
\end{aligned}
\]

\medskip

Each of the above $48$ pairings contributes a term
\[
\lambda_1^4 \times (\text{product of four }\delta\text{'s}),
\]
and these are precisely the nonzero terms in
$\kappa^*_{(ij)(lm)(st)(uv)}$.

We classify the index patterns of $\{i, j, l, m, s, t, u, v \}$ according to the number of distinct values, subject to the conditions
\[
1\le i\le j\le k,\qquad
1\le l\le m\le k,\qquad
1\le s\le t\le k,\qquad
1\le u\le v\le k.
\]

\textbf{Class A: one distinct value.}
All eight indices are identical:
\[
i=j=l=m=s=t=u=v.
\]
Then all $48$ connected pairings contribute, and hence
\[
\kappa^*=48\lambda_1^4.
\]
The number of such patterns is
\[
k.
\]

\textbf{Class B1: two distinct values, two diagonal pairs and two off-diagonal pairs of the same type}
\\
Consider the case $(i, j) = (l, m)= (a,a), (s,t) = (u,v) =(a,b)$, where $a<b$. Then
$i=j=l=m=s=u=a,\ t=v=b.$
We now evaluate all $48$ admissible connected pairings.

\subsection*{Cycle $B_1-B_2-B_4-B_3-B_1$}

\[
\begin{array}{c|c}
\text{Pairing} & \text{Value} \\
\hline
(i,l)(j,s)(m,u)(t,v) & 1\\
(i,l)(j,s)(m,v)(t,u) & 0\\
(i,l)(j,t)(m,u)(s,v) & 0\\
(i,l)(j,t)(m,v)(s,u) & 0\\
(j,l)(i,s)(m,u)(t,v) & 1\\
(j,l)(i,s)(m,v)(t,u) & 0\\
(j,l)(i,t)(m,u)(s,v) & 0\\
(j,l)(i,t)(m,v)(s,u) & 0\\
(i,m)(j,s)(l,u)(t,v) & 1\\
(i,m)(j,s)(l,v)(t,u) & 0\\
(i,m)(j,t)(l,u)(s,v) & 0\\
(i,m)(j,t)(l,v)(s,u) & 0\\
(j,m)(i,s)(l,u)(t,v) & 1\\
(j,m)(i,s)(l,v)(t,u) & 0\\
(j,m)(i,t)(l,u)(s,v) & 0\\
(j,m)(i,t)(l,v)(s,u) & 0
\end{array}
\]

This cycle contributes \(4\).

\subsection*{Cycle $B_1-B_2-B_3-B_4-B_1$}

\[
\begin{array}{c|c}
\text{Pairing} & \text{Value} \\
\hline
(i,l)(j,u)(m,s)(v,t) & 1\\
(i,l)(j,u)(m,t)(v,s) & 0\\
(i,l)(j,v)(m,s)(u,t) & 0\\
(i,l)(j,v)(m,t)(u,s) & 0\\
(j,l)(i,u)(m,s)(v,t) & 1\\
(j,l)(i,u)(m,t)(v,s) & 0\\
(j,l)(i,v)(m,s)(u,t) & 0\\
(j,l)(i,v)(m,t)(u,s) & 0\\
(i,m)(j,u)(l,s)(v,t) & 1\\
(i,m)(j,u)(l,t)(v,s) & 0\\
(i,m)(j,v)(l,s)(u,t) & 0\\
(i,m)(j,v)(l,t)(u,s) & 0\\
(j,m)(i,u)(l,s)(v,t) & 1\\
(j,m)(i,u)(l,t)(v,s) & 0\\
(j,m)(i,v)(l,s)(u,t) & 0\\
(j,m)(i,v)(l,t)(u,s) & 0
\end{array}
\]

This cycle contributes \(4\).

\subsection*{Cycle $B_1-B_3-B_2-B_4-B_1$}

\[
\begin{array}{c|c}
\text{Pairing} & \text{Value} \\
\hline
(i,s)(j,u)(t,l)(m,v) & 0\\
(i,s)(j,u)(t,m)(l,v) & 0\\
(i,s)(j,v)(t,l)(m,u) & 0\\
(i,s)(j,v)(t,m)(l,u) & 0\\
(j,s)(i,u)(t,l)(m,v) & 0\\
(j,s)(i,u)(t,m)(l,v) & 0\\
(j,s)(i,v)(t,l)(m,u) & 0\\
(j,s)(i,v)(t,m)(l,u) & 0\\
(i,s)(j,u)(t,m)(l,v) & 0\\
(i,s)(j,u)(t,l)(m,v) & 0\\
(i,s)(j,v)(t,m)(l,u) & 0\\
(i,s)(j,v)(t,l)(m,u) & 0\\
(j,s)(i,u)(t,m)(l,v) & 0\\
(j,s)(i,u)(t,l)(m,v) & 0\\
(j,s)(i,v)(t,m)(l,u) & 0\\
(j,s)(i,v)(t,l)(m,u) & 0
\end{array}
\]

This cycle contributes \(0\).

Therefore, in total, 
\[
4+4+0=8.
\]
Thus,
\[
\kappa^*_{(a,a)(a,a)(a,b)(a,b)}
=
8\lambda_1^4.
\]
The other cases in Class B1 yield the same value of $\kappa^*$.

For each pair $\{a,b\}$, there are $12$ such patterns. Hence, the total number of
patterns is
\[
12\binom{k}{2}=6k(k-1).
\]

\textbf{Class B2: two distinct values, all four pairs off-diagonal of the same type.}

For $a<b$, 
\[
(i,j)=(l,m)=(s,t)=(u,v)=(a,b).
\]
Thus
\[
i=l=s=u=a,\qquad j=m=t=v=b.
\]
We now evaluate all $48$ admissible connected pairings.
\subsection*{Cycle $B_1-B_2-B_4-B_3-B_1$}

\[
\begin{array}{c|c}
\text{Pairing} & \text{Value} \\
\hline
(i,l)(j,s)(m,u)(t,v) & 0\\
(i,l)(j,s)(m,v)(t,u) & 0\\
(i,l)(j,t)(m,u)(s,v) & 0\\
(i,l)(j,t)(m,v)(s,u) & 1\\
(j,l)(i,s)(m,u)(t,v) & 0\\
(j,l)(i,s)(m,v)(t,u) & 0\\
(j,l)(i,t)(m,u)(s,v) & 0\\
(j,l)(i,t)(m,v)(s,u) & 0\\
(i,m)(j,s)(l,u)(t,v) & 0\\
(i,m)(j,s)(l,v)(t,u) & 0\\
(i,m)(j,t)(l,u)(s,v) & 0\\
(i,m)(j,t)(l,v)(s,u) & 0\\
(j,m)(i,s)(l,u)(t,v) & 1\\
(j,m)(i,s)(l,v)(t,u) & 0\\
(j,m)(i,t)(l,u)(s,v) & 0\\
(j,m)(i,t)(l,v)(s,u) & 0
\end{array}
\]

This cycle contributes \(2\).

\subsection*{Cycle $B_1-B_2-B_3-B_4-B_1$}

\[
\begin{array}{c|c}
\text{Pairing} & \text{Value} \\
\hline
(i,l)(j,u)(m,s)(v,t) & 0\\
(i,l)(j,u)(m,t)(v,s) & 0\\
(i,l)(j,v)(m,s)(u,t) & 0\\
(i,l)(j,v)(m,t)(u,s) & 1\\
(j,l)(i,u)(m,s)(v,t) & 0\\
(j,l)(i,u)(m,t)(v,s) & 0\\
(j,l)(i,v)(m,s)(u,t) & 0\\
(j,l)(i,v)(m,t)(u,s) & 0\\
(i,m)(j,u)(l,s)(v,t) & 0\\
(i,m)(j,u)(l,t)(v,s) & 0\\
(i,m)(j,v)(l,s)(u,t) & 0\\
(i,m)(j,v)(l,t)(u,s) & 0\\
(j,m)(i,u)(l,s)(v,t) & 1\\
(j,m)(i,u)(l,t)(v,s) & 0\\
(j,m)(i,v)(l,s)(u,t) & 0\\
(j,m)(i,v)(l,t)(u,s) & 0
\end{array}
\]

This cycle contributes \(2\).

\subsection*{Cycle $B_1-B_3-B_2-B_4-B_1$}

\[
\begin{array}{c|c}
\text{Pairing} & \text{Value} \\
\hline
(i,s)(j,u)(t,l)(m,v) & 0\\
(i,s)(j,u)(t,m)(l,v) & 0\\
(i,s)(j,v)(t,l)(m,u) & 0\\
(i,s)(j,v)(t,m)(l,u) & 1\\
(j,s)(i,u)(t,l)(m,v) & 0\\
(j,s)(i,u)(t,m)(l,v) & 0\\
(j,s)(i,v)(t,l)(m,u) & 0\\
(j,s)(i,v)(t,m)(l,u) & 0\\
(i,s)(j,u)(t,m)(l,v) & 0\\
(i,s)(j,u)(t,l)(m,v) & 0\\
(i,s)(j,v)(t,m)(l,u) & 1\\
(i,s)(j,v)(t,l)(m,u) & 0\\
(j,s)(i,u)(t,m)(l,v) & 0\\
(j,s)(i,u)(t,l)(m,v) & 0\\
(j,s)(i,v)(t,m)(l,u) & 0\\
(j,s)(i,v)(t,l)(m,u) & 0
\end{array}
\]

This cycle contributes \(2\).

Therefore,
\[
2+2+2=6.
\]
Hence,
\[
\kappa^*_{(a,b)(a,b)(a,b)(a,b)}
=
6\lambda_1^4.
\]
The number of such patterns is
\[
\binom{k}{2}=\frac{k(k-1)}{2}.
\]

\textbf{Class B3: two distinct values, two diagonal pairs of the different type and two off-diagonal pairs.}
\\
For $a<b$, the four blocks consist of one $\{a, a\}$ block, two $\{a, b\}$ blocks, and one $\{b, b\}$ block.
Consider the case
\[
(i,j)=(a,a),\qquad (l,m)=(s,t)=(a,b),\qquad (u,v)=(b,b),
\qquad a<b.
\]
Thus
\[
i=j=l=s=a,\qquad m=t=u=v=b.
\]
We now evaluate all $48$ admissible connected pairings.
\subsection*{Cycle $B_1-B_2-B_4-B_3-B_1$}

\[
\begin{array}{c|c}
\text{Pairing} & \text{Value} \\
\hline
(i,l)(j,s)(m,u)(t,v) & 1\\
(i,l)(j,s)(m,v)(t,u) & 1\\
(i,l)(j,t)(m,u)(s,v) & 0\\
(i,l)(j,t)(m,v)(s,u) & 0\\
(j,l)(i,s)(m,u)(t,v) & 1\\
(j,l)(i,s)(m,v)(t,u) & 1\\
(j,l)(i,t)(m,u)(s,v) & 0\\
(j,l)(i,t)(m,v)(s,u) & 0\\
(i,m)(j,s)(l,u)(t,v) & 0\\
(i,m)(j,s)(l,v)(t,u) & 0\\
(i,m)(j,t)(l,u)(s,v) & 0\\
(i,m)(j,t)(l,v)(s,u) & 0\\
(j,m)(i,s)(l,u)(t,v) & 0\\
(j,m)(i,s)(l,v)(t,u) & 0\\
(j,m)(i,t)(l,u)(s,v) & 0\\
(j,m)(i,t)(l,v)(s,u) & 0
\end{array}
\]

This cycle contributes \(4\).

\subsection*{Cycle $B_1-B_2-B_3-B_4-B_1$}

\[
\begin{array}{c|c}
\text{Pairing} & \text{Value} \\
\hline
(i,l)(j,u)(m,s)(v,t) & 0\\
(i,l)(j,u)(m,t)(v,s) & 0\\
(i,l)(j,v)(m,s)(u,t) & 0\\
(i,l)(j,v)(m,t)(u,s) & 0\\
(j,l)(i,u)(m,s)(v,t) & 0\\
(j,l)(i,u)(m,t)(v,s) & 0\\
(j,l)(i,v)(m,s)(u,t) & 0\\
(j,l)(i,v)(m,t)(u,s) & 0\\
(i,m)(j,u)(l,s)(v,t) & 0\\
(i,m)(j,u)(l,t)(v,s) & 0\\
(i,m)(j,v)(l,s)(u,t) & 0\\
(i,m)(j,v)(l,t)(u,s) & 0\\
(j,m)(i,u)(l,s)(v,t) & 0\\
(j,m)(i,u)(l,t)(v,s) & 0\\
(j,m)(i,v)(l,s)(u,t) & 0\\
(j,m)(i,v)(l,t)(u,s) & 0
\end{array}
\]

This cycle contributes \(0\).

\subsection*{Cycle $B_1-B_3-B_2-B_4-B_1$}

\[
\begin{array}{c|c}
\text{Pairing} & \text{Value} \\
\hline
(i,s)(j,u)(t,l)(m,v) & 0\\
(i,s)(j,u)(t,m)(l,v) & 0\\
(i,s)(j,v)(t,l)(m,u) & 0\\
(i,s)(j,v)(t,m)(l,u) & 0\\
(j,s)(i,u)(t,l)(m,v) & 0\\
(j,s)(i,u)(t,m)(l,v) & 0\\
(j,s)(i,v)(t,l)(m,u) & 0\\
(j,s)(i,v)(t,m)(l,u) & 0\\
(i,s)(j,u)(t,m)(l,v) & 0\\
(i,s)(j,u)(t,l)(m,v) & 0\\
(i,s)(j,v)(t,m)(l,u) & 0\\
(i,s)(j,v)(t,l)(m,u) & 0\\
(j,s)(i,u)(t,m)(l,v) & 0\\
(j,s)(i,u)(t,l)(m,v) & 0\\
(j,s)(i,v)(t,m)(l,u) & 0\\
(j,s)(i,v)(t,l)(m,u) & 0
\end{array}
\]

This cycle contributes \(0\).

Therefore, in total, 
\[
4+0+0=4.
\]

Hence
\[
\kappa^*_{(a,a)(a,b)(a,b)(b,b)}
=
4\lambda_1^4.
\]
The other cases in Class B3 yield the same value of $\kappa^*$.

For each pair $\{a,b\}$, there are $12$ such patterns. Hence, the total number of
patterns is
\[
12\binom{k}{2}=6k(k-1).
\]

\textbf{Class C1: three distinct values, one diagonal pair and three different off-diagonal pairs}
\\
Let $a < b < c$. Consider the case
\[
(i,j)=(a,a),\quad (l,m)=(a,b),\quad (s,t)=(a,c),\quad (u,v)=(b,c),
\]
Thus
\[
i=j=l=s=a,\qquad m=u=b,\qquad t=v=c.
\]
We now evaluate all $48$ admissible connected pairings.
\subsection*{Cycle $B_1-B_2-B_4-B_3-B_1$}

\[
\begin{array}{c|c}
\text{Pairing} & \text{Value} \\
\hline
(i,l)(j,s)(m,u)(t,v) & 1\\
(i,l)(j,s)(m,v)(t,u) & 0\\
(i,l)(j,t)(m,u)(s,v) & 0\\
(i,l)(j,t)(m,v)(s,u) & 0\\
(j,l)(i,s)(m,u)(t,v) & 1\\
(j,l)(i,s)(m,v)(t,u) & 0\\
(j,l)(i,t)(m,u)(s,v) & 0\\
(j,l)(i,t)(m,v)(s,u) & 0\\
(i,m)(j,s)(l,u)(t,v) & 0\\
(i,m)(j,s)(l,v)(t,u) & 0\\
(i,m)(j,t)(l,u)(s,v) & 0\\
(i,m)(j,t)(l,v)(s,u) & 0\\
(j,m)(i,s)(l,u)(t,v) & 0\\
(j,m)(i,s)(l,v)(t,u) & 0\\
(j,m)(i,t)(l,u)(s,v) & 0\\
(j,m)(i,t)(l,v)(s,u) & 0
\end{array}
\]

This cycle contributes \(2\).

\subsection*{Cycle $B_1-B_2-B_3-B_4-B_1$}

\[
\begin{array}{c|c}
\text{Pairing} & \text{Value} \\
\hline
(i,l)(j,u)(m,s)(v,t) & 0\\
(i,l)(j,u)(m,t)(v,s) & 0\\
(i,l)(j,v)(m,s)(u,t) & 0\\
(i,l)(j,v)(m,t)(u,s) & 0\\
(j,l)(i,u)(m,s)(v,t) & 0\\
(j,l)(i,u)(m,t)(v,s) & 0\\
(j,l)(i,v)(m,s)(u,t) & 0\\
(j,l)(i,v)(m,t)(u,s) & 0\\
(i,m)(j,u)(l,s)(v,t) & 0\\
(i,m)(j,u)(l,t)(v,s) & 0\\
(i,m)(j,v)(l,s)(u,t) & 0\\
(i,m)(j,v)(l,t)(u,s) & 0\\
(j,m)(i,u)(l,s)(v,t) & 0\\
(j,m)(i,u)(l,t)(v,s) & 0\\
(j,m)(i,v)(l,s)(u,t) & 0\\
(j,m)(i,v)(l,t)(u,s) & 0
\end{array}
\]

This cycle contributes \(0\).

\subsection*{Cycle $B_1-B_3-B_2-B_4-B_1$}

\[
\begin{array}{c|c}
\text{Pairing} & \text{Value} \\
\hline
(i,s)(j,u)(t,l)(m,v) & 0\\
(i,s)(j,u)(t,m)(l,v) & 0\\
(i,s)(j,v)(t,l)(m,u) & 0\\
(i,s)(j,v)(t,m)(l,u) & 0\\
(j,s)(i,u)(t,l)(m,v) & 0\\
(j,s)(i,u)(t,m)(l,v) & 0\\
(j,s)(i,v)(t,l)(m,u) & 0\\
(j,s)(i,v)(t,m)(l,u) & 0\\
(i,s)(j,u)(t,m)(l,v) & 0\\
(i,s)(j,u)(t,l)(m,v) & 0\\
(i,s)(j,v)(t,m)(l,u) & 0\\
(i,s)(j,v)(t,l)(m,u) & 0\\
(j,s)(i,u)(t,m)(l,v) & 0\\
(j,s)(i,u)(t,l)(m,v) & 0\\
(j,s)(i,v)(t,m)(l,u) & 0\\
(j,s)(i,v)(t,l)(m,u) & 0
\end{array}
\]

This cycle contributes \(0\).

Therefore,
\[
2+0+0=2.
\]

Hence
\[
\kappa^*_{(a,a)(a,b)(a,c)(b,c)}
=
2\lambda_1^4.
\]
The other cases in Class C1 yield the same value of $\kappa^*$.
There are $72$ such patterns for each triple $\{a,b,c\}$; hence, the number of patterns in Class C1 is
\[
72\binom{k}{3}
=
12k(k-1)(k-2).
\]

\textbf{Class C2: three distinct values, two copies of one off-diagonal pair and
two copies of another off-diagonal pair}\\
Let $a < b < c$. Consider the case $(i, j) =(l,m)= (a, b),\ (s,t) = (u, v)=(a,c)$, so that $i=l=s=u=a,\ j=m=b,\ t=v=c.$
\subsection*{Cycle $B_1-B_2-B_4-B_3-B_1$}
\[
\begin{array}{c|c}
\text{Pairing} & \text{Value} \\
\hline
(i,l)(j,s)(m,u)(t,v) & 0\\
(i,l)(j,s)(m,v)(t,u) & 0\\
(i,l)(j,t)(m,u)(s,v) & 0\\
(i,l)(j,t)(m,v)(s,u) & 0\\
(j,l)(i,s)(m,u)(t,v) & 0\\
(j,l)(i,s)(m,v)(t,u) & 0\\
(j,l)(i,t)(m,u)(s,v) & 0\\
(j,l)(i,t)(m,v)(s,u) & 0\\
(i,m)(j,s)(l,u)(t,v) & 0\\
(i,m)(j,s)(l,v)(t,u) & 0\\
(i,m)(j,t)(l,u)(s,v) & 0\\
(i,m)(j,t)(l,v)(s,u) & 0\\
(j,m)(i,s)(l,u)(t,v) & 1\\
(j,m)(i,s)(l,v)(t,u) & 0\\
(j,m)(i,t)(l,u)(s,v) & 0\\
(j,m)(i,t)(l,v)(s,u) & 0
\end{array}
\]

This cycle contributes \(1\).

\subsection*{Cycle $B_1-B_2-B_3-B_4-B_1$}

\[
\begin{array}{c|c}
\text{Pairing} & \text{Value} \\
\hline
(i,l)(j,u)(m,s)(v,t) & 0\\
(i,l)(j,u)(m,t)(v,s) & 0\\
(i,l)(j,v)(m,s)(u,t) & 0\\
(i,l)(j,v)(m,t)(u,s) & 0\\
(j,l)(i,u)(m,s)(v,t) & 0\\
(j,l)(i,u)(m,t)(v,s) & 0\\
(j,l)(i,v)(m,s)(u,t) & 0\\
(j,l)(i,v)(m,t)(u,s) & 0\\
(i,m)(j,u)(l,s)(v,t) & 0\\
(i,m)(j,u)(l,t)(v,s) & 0\\
(i,m)(j,v)(l,s)(u,t) & 0\\
(i,m)(j,v)(l,t)(u,s) & 0\\
(j,m)(i,u)(l,s)(v,t) & 1\\
(j,m)(i,u)(l,t)(v,s) & 0\\
(j,m)(i,v)(l,s)(u,t) & 0\\
(j,m)(i,v)(l,t)(u,s) & 0
\end{array}
\]

This cycle contributes \(1\).

\subsection*{Cycle $B_1-B_3-B_2-B_4-B_1$}

\[
\begin{array}{c|c}
\text{Pairing} & \text{Value} \\
\hline
(i,s)(j,u)(t,l)(m,v) & 0\\
(i,s)(j,u)(t,m)(l,v) & 0\\
(i,s)(j,v)(t,l)(m,u) & 0\\
(i,s)(j,v)(t,m)(l,u) & 0\\
(j,s)(i,u)(t,l)(m,v) & 0\\
(j,s)(i,u)(t,m)(l,v) & 0\\
(j,s)(i,v)(t,l)(m,u) & 0\\
(j,s)(i,v)(t,m)(l,u) & 0\\
(i,s)(j,u)(t,m)(l,v) & 0\\
(i,s)(j,u)(t,l)(m,v) & 0\\
(i,s)(j,v)(t,m)(l,u) & 0\\
(i,s)(j,v)(t,l)(m,u) & 0\\
(j,s)(i,u)(t,m)(l,v) & 0\\
(j,s)(i,u)(t,l)(m,v) & 0\\
(j,s)(i,v)(t,m)(l,u) & 0\\
(j,s)(i,v)(t,l)(m,u) & 0
\end{array}
\]

This cycle contributes \(0\).

Therefore,
\[
1+1+0=2.
\]

Hence,
\[
\kappa^*_{(a,b)(a,b)(a,c)(a,c)}
=
2\lambda_1^4.
\]
The other cases in Class C2 yield the same value of $\kappa^*$.

There are $18$ such patterns for each triple $\{a,b,c\}$.
Thus the total number of Class C2 patterns is
\[
18\binom{k}{3}
=
3k(k-1)(k-2).
\]

\textbf{Class D: four distinct values, 4 different off-diagonal pairs which makes a circle}
\\
Let $a<b<c<d$ and
\[
(i,j) =(a,b),\ (l,m)=(b, c),\ (s,t)=(c, d),\ (u,v)=(a,d).
\]
Thus
\[
i=u=a,\qquad j=l=b,\qquad m=s=c,\qquad t=v=d.
\]

\subsection*{Cycle $B_1-B_2-B_4-B_3-B_1$}

\[
\begin{array}{c|c}
\text{Pairing} & \text{Value} \\
\hline
(i,l)(j,s)(m,u)(t,v) & 0\\
(i,l)(j,s)(m,v)(t,u) & 0\\
(i,l)(j,t)(m,u)(s,v) & 0\\
(i,l)(j,t)(m,v)(s,u) & 0\\
(j,l)(i,s)(m,u)(t,v) & 0\\
(j,l)(i,s)(m,v)(t,u) & 0\\
(j,l)(i,t)(m,u)(s,v) & 0\\
(j,l)(i,t)(m,v)(s,u) & 0\\
(i,m)(j,s)(l,u)(t,v) & 0\\
(i,m)(j,s)(l,v)(t,u) & 0\\
(i,m)(j,t)(l,u)(s,v) & 0\\
(i,m)(j,t)(l,v)(s,u) & 0\\
(j,m)(i,s)(l,u)(t,v) & 0\\
(j,m)(i,s)(l,v)(t,u) & 0\\
(j,m)(i,t)(l,u)(s,v) & 0\\
(j,m)(i,t)(l,v)(s,u) & 0
\end{array}
\]

This cycle contributes \(0\).

\subsection*{Cycle $B_1-B_2-B_3-B_4-B_1$}

\[
\begin{array}{c|c}
\text{Pairing} & \text{Value} \\
\hline
(i,l)(j,u)(m,s)(v,t) & 0\\
(i,l)(j,u)(m,t)(v,s) & 0\\
(i,l)(j,v)(m,s)(u,t) & 0\\
(i,l)(j,v)(m,t)(u,s) & 0\\
(j,l)(i,u)(m,s)(v,t) & 1\\
(j,l)(i,u)(m,t)(v,s) & 0\\
(j,l)(i,v)(m,s)(u,t) & 0\\
(j,l)(i,v)(m,t)(u,s) & 0\\
(i,m)(j,u)(l,s)(v,t) & 0\\
(i,m)(j,u)(l,t)(v,s) & 0\\
(i,m)(j,v)(l,s)(u,t) & 0\\
(i,m)(j,v)(l,t)(u,s) & 0\\
(j,m)(i,u)(l,s)(v,t) & 0\\
(j,m)(i,u)(l,t)(v,s) & 0\\
(j,m)(i,v)(l,s)(u,t) & 0\\
(j,m)(i,v)(l,t)(u,s) & 0
\end{array}
\]

This cycle contributes \(1\).

\subsection*{Cycle $B_1-B_3-B_2-B_4-B_1$}

\[
\begin{array}{c|c}
\text{Pairing} & \text{Value} \\
\hline
(i,s)(j,u)(t,l)(m,v) & 0\\
(i,s)(j,u)(t,m)(l,v) & 0\\
(i,s)(j,v)(t,l)(m,u) & 0\\
(i,s)(j,v)(t,m)(l,u) & 0\\
(j,s)(i,u)(t,l)(m,v) & 0\\
(j,s)(i,u)(t,m)(l,v) & 0\\
(j,s)(i,v)(t,l)(m,u) & 0\\
(j,s)(i,v)(t,m)(l,u) & 0\\
(i,s)(j,u)(t,m)(l,v) & 0\\
(i,s)(j,u)(t,l)(m,v) & 0\\
(i,s)(j,v)(t,m)(l,u) & 0\\
(i,s)(j,v)(t,l)(m,u) & 0\\
(j,s)(i,u)(t,m)(l,v) & 0\\
(j,s)(i,u)(t,l)(m,v) & 0\\
(j,s)(i,v)(t,m)(l,u) & 0\\
(j,s)(i,v)(t,l)(m,u) & 0
\end{array}
\]

This cycle contributes \(0\).

Therefore,
\[
0+1+0=1.
\]

Hence
\[
\kappa^*_{(a,b)(b,c)(c,d)(a,d)}
=
\lambda_1^4.
\]
The other cases in Class D yield the same value of $\kappa^*$.

There are $3$ possible cycles on four labeled blocks and $4!$ ways to assign
the four edges to the four blocks. Hence, there are
\[
3\cdot 4! = 72
\]
patterns for each quadruple $\{a,b,c,d\}$.
The total number of patterns is
\[
72\binom{k}{4}
=
3k(k-1)(k-2)(k-3).
\]

For each class, the value of $\kappa^*_{(ij)(lm)(st)(uv)}$ and the number of patterns are summarized in Table \ref{kappa*_ij_lm_st_uv}.
%
%
%
%
%
\begin{table}[htbp]
\centering
\caption{Values and counts of $\kappa^*_{(ij)(lm)(st)(uv)}$}
\label{kappa*_ij_lm_st_uv}
\begin{tabular}{|c|c|c|}
\hline
Class & Value of $\kappa^*_{(ij)(lm)(st)(uv)}$ & Number of Index Patterns \\
\hline
A & $48\lambda_1^4$ & $k$ \\
\hline
B1 & $8\lambda_1^4$ & $6k(k-1)$ \\
\hline
B2 & $6\lambda_1^4$ & $\displaystyle \frac{k(k-1)}{2}$ \\
\hline
B3 & $4\lambda_1^4$ & $6k(k-1)$ \\
\hline
C1 & $2\lambda_1^4$ & $12k(k-1)(k-2)$ \\
\hline
C2 & $2\lambda_1^4$ & $3k(k-1)(k-2)$ \\
\hline
D & $\lambda_1^4$ & $3k(k-1)(k-2)(k-3)$ \\
\hline
Others & $0$ &
\text{unkown}
 \\
\hline
\end{tabular}
\end{table}
\section{Final Results on Risk}
From the expressions for $\tilde{G}^{-1}$ and $G$, we find that the first-order ($n^{-1}$) term in equation (8) of \cite{Sheena4} is
\begin{equation}
\begin{split}
\mathrm{tr}\Bigl(\tilde{G}^{-1}G \Bigr) &= k\lambda_1^{-1}\lambda_1 + k(2\lambda_1^2)^{-1}(3\lambda_2-\lambda_1^2) + (k(k-1)/2)(\lambda_1^2)^{-1}\lambda_2 \\
&= k + \frac{k}{2}\frac{(\nu-2)^2}{\nu^2}\left(\frac{3\nu^2}{(\nu-2)(\nu-4)}-\frac{\nu^2}{(\nu-2)^2}\right)\\
&\quad+\frac{k(k-1)}{2}\frac{(\nu-2)^2}{\nu^2}\frac{\nu^2}{(\nu-2)(\nu-4)}\\
&=k + \frac{k}{2}\left(\frac{3(\nu-2)}{\nu-4}-1\right)+\frac{k(k-1)}{2}\frac{\nu-2}{\nu-4}\\
&=\frac{k}{2}+\frac{k(k+2)}{2}\frac{\nu-2}{\nu-4}\\
&=p+\frac{k(k+2)}{\nu-4}.
\end{split}
\end{equation}

Consider the second-order terms ($n^{-2}$) in equation (8) of \cite{Sheena4}. Let
$a_i (i=1,\ldots ,10)$ be an index ranging over $R = R_1 \cup R_2$, where
\[
R_1 = \{ i | 1\leq i \leq k \},\qquad R_2= \{(ij) | 1 \leq i \leq j \leq k \}
\]
Then, the three terms of order $n^{-2}$ in equation (8) of \cite{Sheena4} can be rewritten as
\begin{align}
&-8\sum_{a_1,\ldots, a_6 \in R} \tilde{g}^{a_1a_2}\tilde{g}^{a_3 a_4}\tilde{g}^{a_5 a_6} \kappa_{a_2 a_4 a_6 }\kappa^*_{a_1 a_3 a_5} \label{1-1}
\\
&9\sum_{a_1,\ldots, a_{10} \in R}\tilde{g}^{a_1 a_2}\tilde{g}^{a_3 a_4}\tilde{g}^{a_5 a_6}\tilde{g}^{a_7 a_8}\tilde{g}^{a_9 a_{10}}\kappa^*_{a_2 a_3 a_{10}}\kappa^*_{a_5 a_7 a_9}\label{1-2}\\
&\qquad \qquad \times(g_{a_1a_4}g_{a_6 a_8}+g_{a_1a_6}g_{a_4 a_8}+g_{a_1a_8}g_{a_4 a_6}) \nonumber\\
&-3\sum_{a_1, \ldots, a_8 \in R}\tilde{g}^{a_1a_2}\tilde{g}^{a_3 a_4}\tilde{g}^{a_5 a_6}\tilde{g}^{a_7 a_8}\kappa^*_{a_3 a_5 a_7 a_2}
(g_{a_1a_4}g_{a_6 a_8}+g_{a_1a_6}g_{a_4 a_8}+g_{a_1a_8}g_{a_4 a_6}) \label{1-3}
\end{align}

Using the cumulant calculations summarized in Tables \ref{table_kappa_(ij)(lm)(st)}, \ref{table*(ij)(lm)(st)}, \ref{table_kappa_(ij)(lm)st}, and \ref{kappa*_ij_lm_st_uv}, we obtain the following results (the details are omitted):
\begin{align*}
&\eqref{1-1}
=
k\lambda_1^{-4}
\left\{
8\lambda_1^4
+
12(k-1)\lambda_1^2\lambda_2
-
8(k+2)(k+4)\lambda_1\lambda_3
\right\},\\
&\eqref{1-2}
=
k\lambda_1^{-4}
\left\{
18(k+1)\lambda_1^4
-
108\lambda_1^2\lambda_2
+
18(2k^2+9k+16)\lambda_2^2
\right\},\\
&\eqref{1-3}
=
k\lambda_1^{-4}
\left\{
9\lambda_1^4
+
18(k+2)\lambda_1^2\lambda_2
-
9(2k^2+11k+14)\lambda_2^2
\right\}.
\end{align*}

Consequently, the second-order term is given by
\[
\begin{aligned}
&\eqref{1-1}+\eqref{1-2}+\eqref{1-3}\\
&=
k\lambda_1^{-4}
\Big\{
(18k+35)\lambda_1^4
+
6(5k-14)\lambda_1^2\lambda_2
\\
&\qquad\qquad-
8(k+2)(k+4)\lambda_1\lambda_3
+
9(2k^2+7k+18)\lambda_2^2
\Big\}\\
&=
\frac{k}{(\nu-6)(\nu-4)^2}\\
&\times
\Big[
(10k^2+63k+49)\nu^3
\\
&\qquad-(116k^2+858k+590)\nu^2
\\
&\qquad+(344k^2+3276k+1800)\nu
\\
&\qquad-(304k^2+3912k+2192)
\Big].
\end{aligned}
\]
The final form of the right-hand side of \eqref{second-order_approximation} is
\begin{align*}
&\frac{1}{2n}\Big(p+\frac{k(k+2)}{\nu-4}\Big)\\
&
+\frac{k}{24n^2(\nu-6)(\nu-4)^2}\\
&\times
\Big[
(10k^2+63k+49)\nu^3
\\
&\qquad-(116k^2+858k+590)\nu^2
\\
&\qquad+(344k^2+3276k+1800)\nu
\\
&\qquad-(304k^2+3912k+2192)
\Big].
\end{align*}

\bibliographystyle{plain}
\bibliography{toukei}
\end{document}